\documentclass[letterpaper,12pt]{article}

\usepackage{mathptmx}

\usepackage{graphicx}

\usepackage[authoryear,comma,sectionbib]{natbib} 

\usepackage{subfigure}
\usepackage{amsmath}
\usepackage{amsfonts}
\usepackage{multirow}
\usepackage{epsfig,lscape}
\usepackage{epstopdf}
\usepackage{multibib}
\newcites{nma}{References for the MRSA application}

\newcommand{\cip}{\mbox{$\perp\!\!\!\perp$}}

\title{A causal inference approach to network meta-analysis}

\author{M. E. Schnitzer$^{1}$, 
R. J. Steele$^{2}$, M. Bally$^{3}$, and I. Shrier$^{4}$\vspace{0.2in}\\
$^{1}$Universit\'e de Montr\'eal, $^{2}$McGill University\\
$^{3}$Centre de recherche du Centre hospitalier de l'Universit\'e de Montr\'eal\\ 
$^{4}$Centre for Clinical Epidemiology, Lady Davis Institute\\
}

\begin{document}

\maketitle



\label{firstpage}
\begin{abstract}
While standard meta-analysis pools the results from randomized trials that compare two treatments, network meta-analysis aggregates the results of randomized trials comparing a wider variety of treatment options. However, it is unclear whether the aggregation of effect estimates across heterogeneous populations will be consistent for a meaningful parameter when not all treatments are evaluated on each population. Drawing from counterfactual theory and the causal inference framework, we define the population of interest in a network meta-analysis and define the target parameter under a series of nonparametric structural assumptions. This allows us to determine the requirements for identifiability of this parameter, enabling a description of the conditions under which network meta-analysis is appropriate and when it might mislead decision making. We then adapt several modeling strategies from the causal inference literature to obtain consistent estimation of the intervention-specific mean outcome and model-independent contrasts between treatments. Finally, we perform a reanalysis of a systematic review to compare the efficacy of antibiotics on suspected or confirmed methicillin-resistant \emph{Staphylococcus aureus} in hospitalized patients.
\end{abstract}

\section{Introduction}

While individual studies are rarely used to inform scientific or medical decision making~\citep{Slavin:001}, multiple sources of evidence may be aggregated in order to offer more generalizable and precise comparisons between treatments~\citep{Lumley:001,Salanti:001,Caldwell:001,Lu:001}.  Meta-analysis, which is the statistical synthesis of multiple study results, is often considered the highest form of quantitative evidence due to its ability to combine all relevant information in the scientific literature. However, because of such issues as effect heterogeneity across study populations and methodology that does not necessarily account for all sources of bias, the status of meta-analysis as the ``gold standard'' of medical knowledge has been questioned~\citep{Berlin:001}.

Standard meta-analysis compares two treatments of interest (or, for instance, an active treatment and placebo). When many treatments for a common condition are tested and made available over time, the medical literature may then contain multiple randomized controlled trials (RCTs) with various treatment comparisons on potentially different populations. Without additional guidance, clinicians and patients are left to informally synthesize information in the available studies in order to determine an optimal treatment decision. A \emph{network meta-analysis} statistically aggregates the results from the relevant RCTs in order to obtain an estimate of the contrast between each pair of treatments. In particular, this type of analysis can produce estimates of contrasts even when no RCT directly compared the two treatments of interest directly.

Each RCT in the network may be performed on populations that differ in terms of their baseline characteristics. These population-specific variables may affect the average response to treatment so that in order to combine inference involving the means, it might be beneficial to control for such variables~\citep{Salanti:002}. Furthermore, it has been noted that if these characteristics not only differentially affect response to treatment, but also the initial study design choice of which treatments to compare, then these variables may confound the overall effect estimate~\citep{Jansen:001,Berlin:001}. As an example, \citet{Jansen:001} suggest that the baseline severity of patients recruited into a study can be related to the type of treatments investigated in the study and also affect the average outcome at the end of the study. As we demonstrate in this paper, such ``study-level confounding'' must be adjusted for in order to obtain consistent estimation of average treatment effects.

In this paper, we consider the setting where individual patient data are not available so that the observed data is limited to average covariate and outcome values in addition to study-level information (which we refer to as ``aggregate'' or study-level data). We begin by describing past parametric approaches to network meta-analysis where the parameter of interest is dependent on the model specification and where the absence of effect heterogeneity is often required a priori. Using the counterfactual framework, we propose a novel definition of a marginal and model-independent causal parameter of interest in network meta-analysis and delineate the assumptions required to estimate this parameter in the presence of measured study-level confounders. We are then able to clarify conditions under which a network meta-analysis is appropriate and when it might mislead decision making regardless of estimation method used. We describe several marginal estimation methods adapted from the single study causal inference setting, including a doubly robust and semiparametric locally efficient Targeted Maximum Likelihood Estimator, and then compare these methods in a simulation study. Finally, we perform a reanalysis of the systematic review by~\citet{Bally:001} to compare the efficacy of antibiotics on suspected or confirmed methicillin-resistant \emph{Staphylococcus aureus} (MRSA) in hospitalized patients.

\section{The observed data}
Each RCT is assumed to randomly sample subjects from a wider population, called a \emph{superpopulation}. Within the RCT, randomization assigns subjects to two or more groups, each one receiving a treatment. These groups are often referred to as \emph{treatment arms}. Due to randomization and random sampling, each group is a representative sample from the superpopulation. Therefore, each arm can be thought of as a distinct study on the same superpopulation. The superpopulations targeted by the RCTs may differ in terms of their characteristics due to, for example, each trial's physical and temporal location, the individual inclusion and exclusion criteria, and the recruitment sample size targets. Therefore, if effect heterogeneity exists (i.e. if the relative treatment effects at the subject level depend on baseline covariate values), one would not expect the average relative treatment effects to necessarily be equal across superpopulations.

More formally, the superpopulation is the conceptual group of essentially infinite size from which the study sample is selected~\citep{Robins:009}. A measure of some outcome ($Y$) is taken on each subject in the RCT arm.  In this article, we will generally consider the example where the sample mean and standard deviation of $Y$ are the summary statistics computed in each RCT. 

Let $A_{ij}$ be the intervention received by subjects in arm $j$ of a particular RCT indexed by $i$.  For this arm, we observe an estimated mean outcome $\bar{Y}_{ij}$ and standard deviation $S_{ij}$.   Let $O_i=(W_i, n_i, \{N_{ij}, A_{ij}, \bar{Y}_{ij}, S_{ij}\};j=1,...,n_i),i=1,...,N$ where $W_i$ is study baseline information and $n_i$ is the number of arms in the study.  
For the $j-$th arm of RCT $i$, let $N_{ij}$ be the number of subjects and $N$ be the total number of RCTs in the sample.

Because we are interested in summarizing effects across multiple superpopulations, we are arguably attempting to estimate effects in a \emph{metapopulation} that contains the individual superpopulations from each study. For the purpose of this paper, we define the metapopulation as the union of possible study superpopulations and define our parameters of interest with respect to this metapopulation. In particular, we assume that the individual $O_i$ vectors are independently drawn from the metapopulation and identically distributed.  

\section{Past approaches to network meta-analysis}~\label{otherapp}
		

Standard approaches in network meta-analysis where only aggregate data are observed place a hierarchical model on either the study-specific contrasts (e.g. the difference in means, $\bar{Y}_{i1}-\bar{Y}_{i2}$) or the arm-specific outcomes ($\bar{Y}_{ij}$) and specify a within-study correlation structure~\citep{Lu:001,Salanti:001,Dias:001,Zhang:001}. As the absence of effect heterogeneity is often required, a priori~\citep{Cope:001} and post-hoc~\citep{Lu:002} investigation of this assumption is routinely recommended. The reader is referred to published guidance~\citep{Dias:003,Jansen:002} and to an example of how heterogeneity was accounted for in an economic analysis~\citep{Welton:001}. There has been recent heated debate about the appropriateness of arm-based estimation methods~\citep{Dias:001,Hong:001}.

The effect targeted in a hierarchical model depends on the contrast-type chosen and the parametrization of the model, and may or may not correspond to a marginal effect as we define further on. For binary outcomes, due to the non-collapsibility of the logistic regression model \citep{Gail:001} in particular, adjustment for covariates in such a model changes the true value of the ``effect'' parameter being estimated. This type of modeling strategy may therefore be biased for the estimation of a marginal effect. Even in linear models, the inclusion of treatment interactions with covariates can also bias the value of the coefficient of treatment relative to the marginal effect. \citet{Zhang:001} and \citet{Zhang:002} take a missing data perspective and model the arm-specific outcomes using a Bayesian hierarchical model to estimate marginal parameters.  While neither approach has yet been extended to incorporate covariates, the former paper assumes that treatments are applied to studies at-random while the latter allows for estimation in a not-at-random context by explicitly specifying the unobservable selection mechanism. 

While adjustment for covariates is rare in practice, \citet{Jansen:001} introduced the notion of adapting Pearl's causal directed acyclic graphs (DAGs) to this setting~\citep{Pearl:002} in order to assist in covariate selection. As a general rule,~\citet{Jansen:001} advocate for the adjustment of all modifiers of the relative treatment effects across comparisons. They also discourage adjustment for covariates that are not effect modifiers due to the fact that they may \emph{induce} bias in the meta-analysis.

\section{The counterfactual approach}

 Let $Y^a$ be the potential (or counterfactual) outcome of a random subject drawn from the metapopulation had that subject received treatment $A=a$. In an RCT, each study arm produces an estimate of the superpopulation-specific mean of the outcome $Y^a$ under the treatment assigned. Let the true mean of the potential outcome under treatment $a$ for the superpopulation targeted in study $i$ be denoted $M^a_i:=E(Y^a\mid P_i)$ where $P_i$ represents the superpopulation targeted in study $i$. Let $\Sigma^a_i:=\sqrt{Var(Y^a\mid P_i)}$ be the standard deviation of the potential outcomes in $P_i$ under treatment $a$. Now suppose that each superpopulation is independently drawn from a metapopulation, $\mathcal{P}=\bigcup_{i\in \mathfrak{s}_P} P_i$, the union of all possible study superpopulations indexed by the set $\mathfrak{s}_P$. A marginal target parameter in a meta-analysis is $M^a:=E(Y^a)=E(M^a_i)$, which represents the mean outcome under treatment $a$ on the metapopulation. The standard deviation of the overall outcome distribution is $\Sigma^a:=\sqrt{Var(Y^a)}=\sqrt{E \{Var(Y^a\mid P_i)\} + Var\{E(Y^a\mid P_i)\}}=\sqrt{E({\Sigma^a_i}^2) + Var(M^a_i)}$, representing the within and between study heterogeneity in the outcome under treatment. Due to treatment arm randomization and random sampling, $\bar{Y}_{ij}$ is an unbiased estimate of $M^{A_{ij}}_i$, the mean potential outcome under the observed treatment, and $S_{ij}$ is a consistent estimate of $\Sigma^{A_{ij}}_i$, the potential standard deviation under the observed treatment.

	For two treatments, $A=a$ and $b$, with corresponding means $M^a$ and $M^b$, we can define a causal effect as the contrast between the mean outcome when the entire metapopulation is treated according to one treatment versus another. For instance, for binary outcomes we may define the causal risk difference as $M^a-M^b$ and the causal risk ratio as $M^a/M^b$. 
	
The patient sample in any given study arm may not be representative of the metapopulation, for which the effect of interest is defined. In addition, because treatment was not randomly allocated across different RCTs, the collection of mean outcomes observed under a given treatment $a$ may not be representative of the metapopulation under treatment $a$. At the design stage, the decision of which treatments to include as arms within an RCT may be influenced by the characteristics of the superpopulation on which the study is taking place. For instance, consider the example of planning a study for a superpopulation with higher disease severity from~\citet{Jansen:001}. Studies including patients with severe disease  are more likely to include an arm with an aggressive treatment. If this occurs, the mean outcome under the aggressive treatment may be different than in a less severe superpopulation. In this situation, we would say that the treatment-mean outcome relationship is confounded at the study level by severity. 

\subsection{A Causal Directed Acyclic Graph (DAG) for network meta-analysis}\label{dag}

Similar to~\citet{Alonso:001}, we assume that heterogeneity in the different superpopulations targeted in the individual RCTs implies that each RCT estimates a different causal effect. Like~\citet{Zhang:001}, we take an ``arm-based" approach to the problem.  Like ~\citet{Jansen:001}, we draw a causal DAG in order to conceptualize the relationship between treatment, study results, and population-specific characteristics.  We arbitrarily choose to intervene on the arm labeled $j$ in each study.  We write $N_{i}=\{N_{ij},j=1,...,n_i\}$, the vector of sample sizes across arms. We will also define $A_i=\{A_{ij},j=1,...,n_i\}$, the vector of treatment assignments evaluated in study $i$, and $A_{i\backslash j}$ to mean the treatment vector excluding some arm $j$.

\begin{figure}[ht]
\centering
\subfigure[] 
{
    \label{Figure2:a}
		\includegraphics[width=0.47\textwidth]{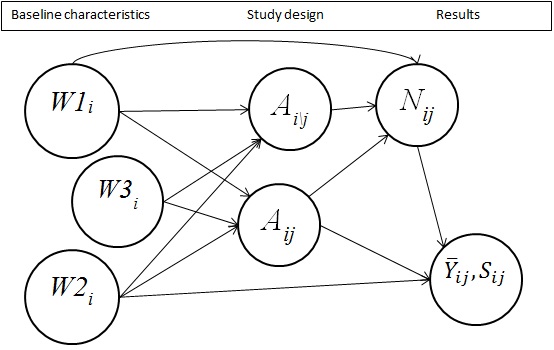}
}
\subfigure[] 
{
    \label{Figure2:b}
		\includegraphics[width=0.47\textwidth]{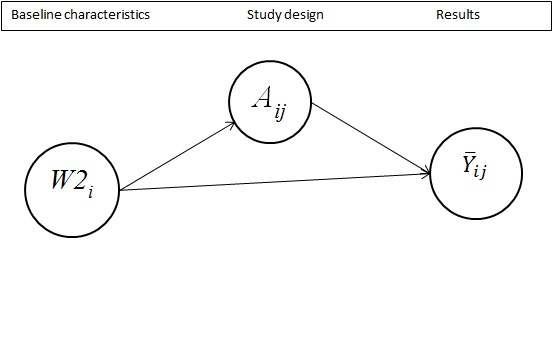}
}
\caption{a) The study-level DAG reflecting the unconfoundedness and time-ordering assumptions made in Sections~\ref{assumptions} and~\ref{Gform} without assuming independence between the sample mean and standard deviation within a study arm. b) The simplified DAG that arises from assuming the independence between the sample mean and standard deviation. Here, $W1_i$, $W2_i$ and $W3_i$ are baseline covariates, $A_{ij}$ and $A_{i\backslash j}$ are the treatments assigned to arms $j$ and the non-$j$ arm(s), respectively, $N_{ij}$ is the sample size of arm $j$, $\bar{Y}_{ij}$ is the mean outcome and $S_{ij}$ is the estimated standard error of the outcome of arm $j$.}\label{Figure2:sub}
\end{figure}

Many of the assumptions presented in detail in Section~\ref{assumptions} are drawn explicitly using the study-level DAGs in Figure~\ref{Figure2:a}. The nodes of the DAG represent variables measured at the level of the RCT and the arrows between them represent the effect of the parent on the child node. For example, the absence of an arrow from $A_{i\backslash j}$ to $\bar{Y}_{ij},S_{ij}$ represents a component of the ``no interference'' assumption that the treatment in one arm will not affect the outcome in another. The arrow from $N_{ij}$ to $\bar{Y}_{ij},S_{ij}$ is present because the sample size within a study arm will affect the distribution of the outcome summary statistics.

The sample size node $N_{ij}$ is determined by the sample size calculation made in the study design phase and also by the success of recruitment. This calculation is inherently conditional on the superpopulation being evaluated, as superpopulation characteristics are taken into account when hypothesizing an effect size and standard error. This calculation is also conditional on the treatments being compared. 

Causal DAGs can be used as a tool to identify which variables must be controlled for in the meta-analysis in order to estimate the treatment-specific metapopulation mean outcome.
Depending on some underlying statistical assumptions that we will investigate in detail in the following sections, these DAGs may simplify to Figure~\ref{Figure2:b}. This happens because we can ignore the mediation path through $N_{ij}$ in order to estimate the total effect of the treatment on outcome. Under these conditions, assuming independence between the variables in $W_i$, the analysis must adjust for all common causes of treatment selection and study outcome distribution. 

Note that the recommendations based on this DAG differ from those of ~\citet{Jansen:001}, who say that the analysis must adjust exclusively for effect modifiers. The assumptions that we list in Section~\ref{assumptions} are explicitly required in the steps we take in Section~\ref{Gform} in order to obtain identifiability of the meta-analysis parameter of interest. 


\subsection{The G-formula and nonparametric identifiability}~\label{Gform}

Suppose we observe the aggregate data $O_i,i=1,...,N$, independently drawn and identically distributed. Using the nonparametric structural equation modeling (NPSEM) of~\citet{Pearl:002}, the metapopulation mean outcome, $M^a$, can be shown to be identifiable (that is, known with infinite data) under the several conditions outlined and discussed in Section~\ref{assumptions}. 

\subsubsection{The observed data generation}~\label{NPSEM}

At the study design stage for RCT $i$, the superpopulation $P_i$ is randomly drawn from the metapopulation $\mathcal{P}$. The selection of $P_i$ determines the population-level covariates $W_i$. The number of study arms $n_i$ and the treatments compared in the study, the multivariate $A_i=\{A_{ij},j=1,...,n_i\}$, are drawn conditional on $W_i$. 
The sample size calculation is carried out based on the choice of treatment comparison and on the sub-population characteristics (i.e. based on the expected effect and precision in that sub-population). This calculation is approximate and the resulting sample size also depends on the success of recruitment. Therefore, the sample sizes for the treatment arms, $N_{i}$, are not deterministic, but are drawn conditional on $A_i$, $n_i$ and $W_i$.
 
The second stage operates at the individual level once subjects are recruited and randomly assigned treatment. Suppose each subject $k$ in arm $j$ of study $i$ has continuous outcome $Y_{ijk},k=1,...,N_{ij}$ (under treatment $A_{ij}$). Each $Y_{ijk}$ is independently drawn from a distribution with mean $M_i^{A_{ij}}$ and standard deviation $\Sigma^{A_{ij}}_i$. The empirical mean outcome in arm $j$ of study $i$ is therefore $\bar{Y}_{ij}=1/N_{ij}\sum_k{Y_{ijk}}$. The standard deviation is estimated as $S^2_{ij}=1/{(N_{ij}-1)}\sum_k (Y_{ijk}-\bar{Y}_{ij})^2$. In addition, subject recruitment yields summary characteristics of the superpopulation, which we assume to include complete information about the covariates $W_i$ that were known at study conception and contributed to the treatment choice. We assume in the following that we do not observe the subject-level data.

Let $\omega_{ij}$ represent the set of estimated summary statistics of the outcome variable from study $i$ arm $j$. For instance, we might have that $\omega_{ij}=\{\bar{Y}_{ij},S_{ij}\}$. Correspondingly, let $\omega_{ij}^a$ be the set of estimates of the counterfactual summary statistics that would arise had arm $j$ been assigned treatment $a$.

Assuming no interference between arms and that the distribution of $\omega_{ij}$ in one arm of a study is conditionally independent of the outcomes in the others and also independent of the total number of arms, the NPSEM that we assume can then be written as
\begin{align*}
&W_i=f_W(\epsilon_W)\\
&n_i=f_n(W_i,\epsilon_n)\\
&A_{i}=f_A(n_i,W_i,\epsilon_A)\\
&N_{i}=f_N(A_{i},n_i,W_i,\epsilon_N)\text{, for }j=1,...,n_i\\
&\omega_{ij}=f_{\omega}(N_{ij},A_{ij},W_i,\epsilon_{\omega})\text{, for }j=1,...,n_i\\
\end{align*}

The probability density function $f(O_{i})$ arising from the NPSEM without intervention can be decomposed as
\begin{align*}
f(O_{i})=& Q_W(W_i)Q_{n}(n_i\mid W_i)g_A(A_{i}\mid n_i,W_i) \times\\
&Q_N(N_{i}\mid A_{i},n_i,W_i)\prod_{j=1}^{n_i}  Q_{\omega}(\omega_{ij}\mid N_{ij},A_{ij},W_{i})
\end{align*}
where $Q_W(W_i)$ corresponds to the density function for $W_i$, $Q_n(n_i\mid W_i)$ corresponds to the density function for $n_i$ conditional on $W_i$, and $g_A(A_i\mid n_i,W_i)$ corresponds to the conditional density function for $A_i$. Within each RCT, $Q_N(N_{i}\mid A_{ij},A_{i\backslash j},n_i,W_i)$ corresponds to the conditional density function for $N_{i}$ and $Q_{\omega}(\omega_{ij}\mid N_{ij},A_{ij},W_{i})$ is the conditional (joint) density for the measured summary statistic(s) in arm $j$. 

\subsubsection{The counterfactual distribution}

Define an intervention as the assignment of treatment strategy $a$ to an arbitrary arm in each study. In other words, for all $i$ we set $A_{ij}=a$ for a single arbitrary arm $j$. The remaining non-$j$ arms receive potential treatments $A^a_{i\backslash j}$. The joint density for the counterfactual data $O^a_{i}=(W_i,n_i,A^a_{i\backslash j},\{\omega_{ij^*}^a,N_{ij^*}^a;j^*=1,...,n_i\})$ can be obtained through the G-formula~\citep{Robins:002}. This joint density function can be written as
\begin{align*}
f(O^a_{i})=& Q_W(W_i)Q_{n}(n_i\mid W_i)g_{A\backslash j}(A^a_{i\backslash j}\mid n_i,W_i)Q_{N}(N_{i}^a\mid A^a_{i\backslash j},n_i,W_i)Q_{\omega}(\omega^a_{ij}\mid N^a_{ij},W_{i})\times\\
&
\prod_{j^*\neq j} Q_{\omega}(\omega^a_{ij^*}\mid N^a_{ij^*},A^a_{ij^*},W_{i})Q_N(N_{ij^*}^a\mid A^a_{i\backslash j},n_i,W_i)
\end{align*}
where $g_{A\backslash j}(A^a_{i\backslash j}\mid n_i,W_i)$ is defined as the conditional (joint) density of the treatments assigned to non-$j$ arms.

\subsubsection{Identifiability for conditionally independent $\bar{Y}$ and $S$}\label{ident_ind}

Suppose we have that $\omega_{ij}=\{\bar{Y}_{ij},S_{ij}\}$, meaning that each study reported the sample means and sample standard deviations of a continuous outcome. We then make the key structural assumption that $\bar{Y}^a_{ij}\cip S^a_{ij} \mid N^a_{ij},W_{i}$ where $N^a_{ij}$ is the counterfactual sample size in study $i$ arm $j$. Let $Y^a_{ijk}$ represent an individual recruited into study $i$ arm $j$ in the counterfactual scenario. The independence assumption arises naturally from the distributional assumption that $Y^a_{ijk} \sim \mathcal{N}(M^a_i,(\Sigma^a_i)^2)$ because $\bar{Y}^a_{ij}$ and $S^a_{ij}$ are the sample mean and standard deviation in superpopulation $P_i$ when $a$ is the treatment assigned. Asymptotically, we have that $\bar{Y}^a_{ij}$ and $S^a_{ij}$ are independent normal variables when the subject-level outcomes are drawn from a distribution with zero skew, such that $E\{(Y^a_{ijk})^3\}=0$~\citep[p. 46]{Ferguson}.
We show in Appendix~\ref{App1} that under this assumption $f(O^a_{i})$ can be decomposed in such a way that the mean outcome, $E(\bar{Y}^a_{ij})=M^a$, can be written independently of the non-$j$ arms, resulting in the simple equality $M^a=\int_W E(\bar{Y}^a_{ij}\mid N^a_{ij},W_i)Q_W(W_i)dW$. Under the unconfoundedness assumption $\bar{Y}^a_{ij}\cip A_{ij}=a\mid N^a_{ij},W_i$, and under the consistency assumption (see next section) we may write the G-formula~\citep{Robins:002} $M^a=\int_{W_i} E(\bar{Y}_{ij}\mid W_i,A_{ij}=a)Q_W(W_i)dW_i$. Therefore, this quantity is identifiable from the data.

Identifiability without assuming this structural independence is possible, and we describe the additional causal assumptions required for this setting in Appendix~\ref{App2}. 

\subsubsection{Identifiability for binary outcomes}
If the original study outcomes are binary (such that $Y_{ijk}=\{0,1\}$), the study means $\bar{Y}_{ij}$ are the proportions of subjects with the indicated outcome. Therefore, $N^a_{ij}\bar{Y}^a_{ij}$ has a binomial distribution with true probability of outcome $M_i^a=E(Y^a\mid P_i)$. Then, $\Sigma^a_i=\sqrt{Var(Y^a\mid P_i)}=\sqrt{M_i^a(1-M_i^a)}$. Similarly, the study arm estimate of the standard deviation is $S_{ij}^a=\sqrt{\bar{Y}^a_{ij}(1-\bar{Y}^a_{ij})}$. In this case, the likelihood will not include a component for $S_{ij}$ so no independence assumption is necessary. The resulting G-formula is still $M^a=\int_{W_i} E(\bar{Y}_{ij}\mid N_{ij},W_i,A_{ij}=a)Q_W(W_i)dW_i$ and will rely on the same unconfoundedness assumption that $\bar{Y}_{ij}^a\cip A_{ij}=a\mid N_{ij},W_i$.

	\subsection{Assumptions}\label{assumptions}
For convenience, here we list the assumptions needed for the identification of $M^a$, corresponding with the NPSEM in Section~\ref{NPSEM} and the DAGs in Figure~\ref{Figure2:sub}. 
 We also comment on the meaning and plausibility of these assumptions in the hypothetical situation where each individual RCT has full compliance. Under full compliance, each RCT arm produces a consistent estimate of the mean outcome in the superpopulation under full adherence to the assigned treatment. 
	
\emph{No interference}. The use of the above counterfactual notation presupposes that the treatment assigned to one study does not affect the counterfactual outcome of another study~\citep{Rubin:003}. A secondary level of interference within an individual study involves the treatment in one study \emph{arm} affecting the outcomes in another study arm. This means that the estimates $\bar{Y}^a_{ij}$ and $S^a_{ij}$ do not depend on the treatment received by another arm of the same RCT. The assumption of no interference will generally not hold for certain studies of infectious disease. For example,  an effective vaccine in one arm may impact the outcome of an unvaccinated subject in the control arm, because the unvaccinated subject will be less likely to be exposed to the disease through herd immunity.

\emph{Unconfoundedness}. (Weak) unconfoundedness~\citep{Imbens:001} is required for the identification of $M^a$. In this context, unconfoundedness is the assumption that the counterfactual sample means under a treatment $a$ are independent of the true treatment received conditional on measured covariates. Specifically, this means that $\bar{Y}_{ij}^a \cip A_{ij}=a \mid N^a_{ij},W_i$. 
In the example DAG of Figure~\ref{Figure2:a}, this corresponds to measuring all the components of node $W2_i$. The validity of this assumption is entirely dependent on the subject-matter, how RCTs in the field are designed, and on the information reported in the RCTs.

\emph{Consistency}. The consistency assumption in this context states that the counterfactual mean of a study arm under a given treatment is the same as the observed result. With notation, this is equivalent to stating that  $\bar{Y}^a_{ij}=\bar{Y}_{ij}$  when $A_{ij}=a$. 
%
Having different definitions of treatment across studies may violate this assumption if all are categorized under the same treatment type and this variation has an impact on the outcome~\citep{Cole:003}. For example, there may be different drug dosages and lengths of follow-up across studies. Disregarding these differences will violate consistency if the various treatment-types have differential effects on the patient outcomes. With some additional unconfoundedness requirements, one might surmount this obstacle using the approach described in \citet{VanderWeele:002}. (Note that this definition of consistency corresponds with the causal assumption and is distinct from the network meta-analysis meaning of the term in e.g.~\citealt{Lu:002}.)

\emph{Positivity}. Finally, we need to evaluate both theoretical and practical positivity. Theoretical positivity is the assumption that, \emph{conditional only on variables required for unconfoundedness}, all studies had a positive probability of being assigned each treatment under investigation. Practical positivity is the condition that for every level of the characteristics $W_i$, there is an \emph{estimated} positive probability of receiving treatment. 

It is important to note that treatment comparisons are based on the same $\mathcal{P}$ and that the target parameter $M^a=E(M^a_i)$ relies on the definition of this metapopulation. If positivity does not hold on some subpopulations it would be necessary to exclude all studies (and corresponding superpopulations) that contain such subpopulations.

%

It is furthermore important to note that the positivity assumption is not the same as requiring that all studies could have realistically been assigned each treatment. In particular, certain treatments may not have been available when some older trials were carried out. If year of study is not required to unconfound the analysis, then the \emph{unconditional} probability may still be non-zero.

\section{Estimation of the treatment-specific metapopulation mean outcome}\label{methods}

\subsection{G-Computation}\label{Gcomp1}

G-Computation procedures based on the G-formula in Section~\ref{Gform} can be used to estimate the target parameter. Here we define a simple procedure resulting from the data requirement that the sample mean and standard deviation are independent within a study arm.  This procedure allows for simple frequentist estimation of the mean effect of treatment. 

This procedure requires estimates for the conditional expectation $E(\bar{Y}_{ij}\mid  W_i,N_{ij}, A_{ij}=a)$ for a given value of treatment. First we must note that while the conditional mean of $\bar{Y}_{ij}$ is independent of $N_{ij}$, its distribution is not. In particular, we have that 
\[Var(\bar{Y}_{ij}\mid W_i, N_{ij},A_{ij})=\frac{1}{N_{ij}}Var(Y_{ijk}\mid W_i, N_{ij},A_{ij})=\frac{1}{N_{ij}}(\Sigma^{A_{ij}}_i)^2.\]
Because $S^2_{ij}$ is a consistent estimate of the superpopulation-level variance under treatment $A_{ij}$, we are able to estimate this variance.

A model for the regression on $\bar{Y}_{ij}$ may be fit by pooling over all arms regardless of treatment assignment. In order to obtain the Best Linear Unbiased Estimator, we can weight by $N_{ij}/S^2_{ij}$. Using this model fit, we predict $\hat{\bar{Y}}^a_{i}=\hat{E}(\bar{Y}_{ij}\mid W_i, A_{ij}=a)$, i.e. the predicted mean under treatment $a$ for each study. The G-Computation estimate is then $\hat{M}_{GCOMP}^a=1/N\sum_{i=1}^N\hat{\bar{Y}}^a_{i}$.

The standard error for the G-Computation estimate is usually computed through nonparametric bootstrap methods~\citep{Snowden:001}. Bootstrap resampling must be done by resampling studies, rather than arms, similar to what is done in a study with clustering~\citep{Efron:001}.

\subsection{Inverse probability of treatment weighting}\label{IPTW}
Likelihood methods, such as G-Computation, require correct parametric specification of the outcome model, which may be difficult to specify. An alternative approach is to utilize propensity score methods, which require the estimation of a model for the treatment received by the arm. For a given treatment type $a$, let $g_a(W_i)$ be an estimate of the probability $P(a\in A_i\mid W_i)$, called the generalized propensity score~\citep{Imbens:001}. 

Despite the small sample size in standard network meta-analysis, one might attempt inverse probability of treatment weighting (IPTW) for the estimation of the marginal parameter. Let $\bar{Y}_i^a$ represent the observed outcome of the arm of study $i$ that received treatment $a$ (or N/A if no arm of study $i$ received treatment $a$). An IPTW estimator for multiple treatments~\citep{Imbens:001} can be represented as 
\begin{equation*}
\hat{M}_{IPTW}^a=1/N  \sum_{i=1}^{N}\frac{\mathbb{I}(a\in A_i)\bar{Y}^a_{i}}{g_a(W_i)}.
\end{equation*} 
Intuitively, this estimator takes a mean of $\bar{Y}_{ij}$ with only the arms treated according to $A_{ij}=a$. It then adjusts this estimate to remove the confounding bias caused by the baseline variables. 

The consistency of this estimator can be shown as follows.
\begin{align*}
\hat{M}_{IPTW}^a&\stackrel{P}{\longrightarrow}E\left[\frac{\bar{Y}^a_{i}\mathbb{I}(a \in A_i)}{P(a\in A_i\mid W_i)}\right]=E\left\{\bar{Y}^a_{i}E\left[\left.\frac{\mathbb{I}(a \in A_i)}{P(a\in A_i\mid W_i)}\right|\bar{Y}^a_i,W_i\right]\right\}=E\left(\bar{Y}^a_{i}\right)=M^a.
\end{align*}

\subsection{Targeted Minimum Loss-based Estimation}\label{TMLE}
Targeted Minimum Loss-based Estimation (TMLE)~\citep{vdl:001,vdl:006} is a framework for the construction of semi-parametric estimators generally applied to the estimation of causal quantities. 
The TMLE procedure is carried out by first fitting a model for the expected value of the arm-based means, $E(\bar{Y}_{ij}\mid W_i, A_{ij}=a)$ which, under the causal assumptions, can equivalently be written as the expectation of the potential outcome had the study evaluated treatment $a$, $E(\bar{Y}_i^a\mid W_i, a\in A_i)$. As in the G-Computation procedure, this model can be estimated by weighing each observation by $N_{ij}/S_{ij}^2$. For each arm in the study, we use this model to obtain $\hat{\bar{Y}}^a_{i}$, predictions of the sample mean of each trial $i$ under treatment $a$. These predictions are then updated by fitting a no-intercept logistic regression using study arms that evaluated treatment $a$. This logistic regression is fit with outcome $\bar{Y}_{ij}$, offset $logit(\hat{\bar{Y}}^a_{i})$, and single covariate $g_a^{-1}(W_i)$, corresponding with the inverse probability weights. Denote the estimate of the coefficient from this regression as $\hat{\epsilon}$. The updated predictions are then $logit(\hat{\bar{Y}}^{a,*}_{i})=logit(\hat{\bar{Y}}^a_{i})+\hat{\epsilon}/g_a(W_i)$, which is calculated for each study. The final targeted estimate for $M^a$ is $\hat{M}_{TMLE}^a=1/N\sum_{i=1}^N \hat{\bar{Y}}^{a,*}_{i}$.
 Note that in order to perform the update step, the means and outcome must be transformed to (0,1) and then subsequently transformed back to the original scale~\citep{Gruber:001}. This can be done using real or empirical bounds. 

This TMLE is consistent under correct specification of the propensity score model or the model for the expected value of the mean outcome (the property of \emph{double robustness}). If both of these models are correct, then TMLE is asymptotically efficient in the class of regular, asymptotically linear estimators in the semiparametric model space~\citep{vdl:006}. More details and a proof of consistency are included in Appendix~\ref{AppTMLE}. 

\section{Simulation study}

In this section we demonstrate that we can obtain consistent estimation of the target parameter $M^a=E(Y^a)$ under the NPSEM using the proposed estimators. We also compare the efficiency of each approach.

While the proposed estimators do not restrict the number of study arms, we fix all simulated studies to have exactly two treatment arms for simplicity. We are interested in estimating the mean outcome of the metapopulation under treatment for each of four treatments of interest. For each study $i=1,...,N$, we generate the population average characteristic, $W_i$ from a Poisson distribution with mean 2. The probabilities of receiving a given treatment are calculated conditional on the value of $W_i$. Two treatment options $A_i$ are then sampled without replacement using the calculated probabilities. Treatments 2 and 4 are generated to be less likely to be chosen with larger $W_i$. The sample size $N_i$ (which we allowed to be common to both arms in the study) is drawn from a Poisson distribution with mean linear in $W_i$ and $A_i$. For each subject within each arm, we draw a baseline covariate $X_{ijk}$ from a Gaussian distribution with mean $W_i$ and constant variance. We set $\beta=(0.8,0.2,1,-0.05)$ to be the treatment-specific coefficients. Outcome values $Y_{ijk}$ are drawn from a Gaussian: $Y_{ijk}\sim\mathcal{N}(X_{ijk}+\beta[A_{ij}],1)$. A summary of the data-generation is presented in Table~\ref{simtable}.

\begin{table}[]
\centering
\caption{Simulation study: data generation}
\label{simtable}
\begin{tabular}{ll}
\hline
Variable                                                                    & Study design: for each $i=1,...,N$                                                                                                                                                                                                                      \\ \hline \\
Number of arms                                                              & $n_i=2$                                                                                                                                                                                                                                                 \\
Study-level covariate                                                       & $W_i\sim Poisson(\mu=2)$                                                                                                                                                                                                                                    \\
Treatments                                                                  & \begin{tabular}[c]{@{}l@{}}$A_i=(A_{i1},A_{i2})$ sampled without replacement with probabilities\\ \quad $p_1 = logit^{-1}(0.4W_i)$\\ \quad $p_2 = logit^{-1}(-0.4W_i)$\\ \quad $p_3 = logit^{-1}(0.8W_i)$\\ \quad $p_4 = logit^{-1}(-0.8W_i)$\end{tabular} \\
\begin{tabular}[c]{@{}l@{}}Sample size \\ (study recruitement)\end{tabular} & \begin{tabular}[c]{@{}l@{}}$N_i\sim Poisson(\mu=5000exp(-0.4W-sum(\gamma[A_{ij}])))$\\ where $\gamma=(-1.5,1,-1,1)$\end{tabular}                                                                                                                                   \\
\\ \hline
                                                                            & Within-study: for each $j=1,2$, $k=1,...,N_i$                                                                                                                                                                                              \\ 
																																						\hline
																																						\\
Subject-level covariate                                                     & $X_{ijk}\sim N(\mu=W_i,\sigma^2=4)$                                                                                                                                                                                                                      \\
Subject-level outcome                                                       & \begin{tabular}[c]{@{}l@{}}$Y_{ijk}\sim N(\mu=X_{ijk}+\beta[A_{ij}],\sigma^2=1)$\\ where $\beta=(0.8,0.2,1,-0.5)$\end{tabular}                                                                                                                                       \\ 
\\
\hline

                                                                            & Observed data: for each $i=1,...,N$, $j=1,2$                                                                                                                                                                                                            \\ \hline \\
Study-level information                                                    & \begin{tabular}[c]{@{}l@{}}$W_i$, $A_i$, and $\bar{Y}_{ij}$ where\\ $\bar{Y}_{ij}=1/N_i\sum_{k=1}^{N_i}Y_{ijk}$
																																							\end{tabular} 
						\\		\\		\hline
\end{tabular}
\end{table}

The sample statistics from each study arm are calculated by taking the mean and standard deviation of $Y_{ijk}$ within each arm. The true treatment-specific superpopulation means are $M^1=2.80,M^2=2.20,M^3=3.00,M^4=1.95$. We are interested in estimating a subset of the contrasts between the treatments, specifically marginal mean differences $M^2-M^1=-0.60$, $M^3-M^1=0.20$, and $M^4-M^1=-0.85$. Note that random effects were not generated in this simple simulation study.

We tested the three methods described in the text (G-Computation, IPTW and TMLE) for $N=15$ and $50$ simulated studies. We used logistic regression models conditional on the covariate for the generalized propensity score for IPTW and TMLE. We ran two scenarios: incorrect and correct outcome model specification. For the correct scenario, linear regression models for the outcome adjusting for treatment type and covariate were used in G-Computation and TMLE. For the incorrect scenario, the outcome was scaled to $(0,1)$ and logistic regression models were used.  We also display results for an unadjusted estimator that merely takes the mean difference in treatment-specific outcomes when available. Variance and confidence intervals were estimated using the nonparametric cluster bootstrap~\citep{Efron:001} where study is considered the cluster (and arms are the individual observations). In Table~\ref{simstudy1}, we present statistics describing the quality of the estimation of all contrasts with treatment 1. These statistics are the percent finite sample bias (``$\%$ Bias''), the standard deviation of the estimates over the simulated data (``SE-MC''), the bootstrap-estimated standard error (``SE-BS''), and the percentage of the 95\% confidence intervals that capture the true effect size (``$\%$ Cov''). Bootstrap resamples that did not allow for an estimate of the contrast (i.e. if either of the treatments did not appear in the resampled data set) were discarded, potentially biasing this standard error estimate.

The unadjusted estimator was greatly biased for the first and third contrasts, indicating that those two contrasts were highly confounded by the simulated study-level covariate. The correctly specified G-Computation estimator had the lowest bias throughout, the smallest standard errors, and near optimal confidence interval coverage. This is to be expected as G-Computation is a function of maximum likelihood parameter estimates with correct parametric specification of the necessary component of the likelihood (namely, the conditional mean of the outcome). However, with an incorrectly specified outcome model, the estimator was biased which caused the coverage to suffer for the third contrast.

IPTW was the most biased estimator and also had the largest variance. The bias largely dissipated when the sample size was increased to $N=50$ studies. IPTW had good coverage except for the third treatment contrast where treatment 4 was rare. The slower convergence of IPTW in the contrast involving treatment 4 can be explained by a higher variance of the estimated weights for that treatment compared to the others. The performance of IPTW has previously been seen to suffer when data support for certain exposure levels is sparse (i.e. under near practical positivity violations)~\citep{Gruber:001}. Truncation of the propensity score at 5\% and 10\% respectively (that is, replacing the bottom $p\%$ of the propensity score with the $p$th percentile) \citep{Cole:002} increases the bias for the first and third contrasts while reducing the variance, with no effect on the coverage (results not shown).

TMLE with correct outcome model specification had bias comparable to G-Computation but slightly higher for $N=15$. For $N=50$, the standard error of TMLE was comparable to that of G-Computation but for $N=15$ it was up to 80 times larger. Regardless, correctly specified TMLE had good coverage throughout. Notably, the bootstrap standard error estimates were comparable to the Monte-Carlo standard error for $N=50$ but diverged for IPTW and TMLE when $N=15$. Certain implementations of TMLE are more sensitive to near practical positivity violations~\citep{Gruber:001,Porter:001,Schnitzer:001}, hence the need for the robust version that involves the logistic regression for the update of the predictions (as described in \citealt{Gruber:001} and for our specific setting in Section~\ref{TMLE}). When the outcome model was misspecified, TMLE also accrued bias for the first and third contrasts, with magnitude comparable to the misspecified G-Computation.  This bias decreased with more studies due to the double robustness of TMLE (making this estimator consistent even when the outcome model is misspecified). Coverage only suffered for the third contrast which was the most biased.

\begin{table}[ht]
  \begin{center}
  \caption{Simulation: Quality of treatment contrast estimation with a Gaussian outcome  (two-arm studies, 1000 simulated datasets).}\label{simstudy1} 
\footnotesize
\begin{tabular}
[htbp]{r|r |c c|  c c|cc}
  \hline
	
		&& \multicolumn{2}{|c|}{$\%$ Bias}& \multicolumn{2}{|c|}{ SE-MC }& \multicolumn{2}{|c}{SE-BS($\%$ Cov)}\\
		&$N$& 15&  50 & 15&50&15& 50 \\
	\hline
	\multicolumn{8}{c}{\emph{Correctly specified models}}\\
	\hline
		\multirow{4}{*}{$M^2-M^1=-0.6$}
		
		&G-Comp   &0  &0& 0.04 &0.02&0.04(91)&0.02(94)\\
		
		&IPTW   & 40 &9& 0.57 &0.46&0.61(89)&0.41(92)\\
		&TMLE   & 4 &0& 0.27 &0.03&0.44(96)&0.05(92)\\
    \hline
		
		\multirow{4}{*}{$M^3-M^1=0.20$}
		
		&G-Comp   & 1 &0& 0.04 &0.02&0.04(91)&0.02(95)\\
		
		&IPTW   & -2 &0& 0.10 &0.04&0.25(99)&0.05(97)\\ 
		&TMLE   & -1 &-1& 0.17 &0.04&0.29(96)&0.04(93)\\
   
\hline
		
		\multirow{4}{*}{$M^4-M^1=-0.85$}

		&G-Comp   & 0 &0& 0.04 &0.02&0.04(89)&0.02(93)\\
		
		&IPTW   & 81 &3& 0.76 &0.74&0.62(63)&0.80(68)\\
		&TMLE   & -9 &0& 0.81 &0.11&0.74(94)&0.21(95)\\
		\hline
   \multicolumn{8}{c}{\emph{Misspecified outcome model}}\\
	\hline
	\multirow{4}{*}{$M^2-M^1=-0.6$}
	& No adjustment   & 101 &103& 0.65 &0.35&0.61(75)&0.34(52)\\
		& G-Comp   & 2&12&0.20&0.13&0.24(98)&0.11(94)\\
		&TMLE   &-8&-2&0.33&0.09&0.46(97)&0.11(96)\\
		 \hline
		\multirow{4}{*}{$M^3-M^1=0.20$}
		&No adjustment   & 5 &-7& 0.37 &0.20&0.38(92)&0.20(93)\\
	& G-Comp   &-1 &7&0.20&0.12&0.18(99)&0.11(95)\\
		&TMLE   &0&0&0.15&0.05&0.28(99)&0.05(96)\\
		 \hline
		\multirow{4}{*}{$M^4-M^1=-0.85$}
				&No adjustment   & 126 &125& 0.69 &0.38&0.61(51)&0.36(18)\\
		& G-Comp   &36 &33&0.53&0.29&0.48(88)&0.24(80)\\
		&TMLE   &44&-24&0.86&0.38&0.75(87)&0.36(75)\\
		
		\hline
    \end{tabular}\end{center}
		\end{table}
		
\section{Application: Antibiotic use on methicillin-resistant \emph{Staphylococcus aureus} infection}\label{application}

We illustrate this causal inference approach and the adapted estimation methods in network meta-analysis with an example from infectious disease research.  An increase in MRSA has spurred investigation of comparative efficacy of different antibiotic treatment options. While the antibiotic vancomycin has been the standard treatment for decades, treatment failures have been noted in patients with serious infections~\citep{Liu:001}. Interest therefore lies in whether alternative antibiotics are as effective as the standard. 
\citet{Bally:001} performed a systematic review and Bayesian network meta-analyses of RCTs of parenteral antibiotics used for treating hospitalized adults with complicated skin and soft-tissue infections (cSSTIs) and hospital-acquired  or ventilator-associated pneumonia.   

We consider the target population of interest to be the population of clinical trial participants with suspected or confirmed MRSA cSSTIs or pneumonia, with corresponding studies published until May 2012. The site of infection and confirmation of MRSA represent important differences in the entrance criteria of the various studies. 24 studies were found. Patients were randomized based on suspicion of  MRSA in all but three studies for which the protocol specified confirmation of presence of MRSA at baseline. 14 studies enrolled subjects with cSSTIs, 7 studies enrolled subjects with hospital-acquired  or ventilator-associated pneumonia, and 3 studies allowed for either indication. The original network meta-analysis of \citet{Bally:001} analyzed each infection site in separate analyses and therefore obtained stratified estimates. Based on the theory we developed, we can account for the potentially different treatment effects in each subpopulation by controlling for subpopulation type as a covariate in the analysis. By doing so, we ask a higher-level yet still clinically interesting question: ``Are the alternative therapies as effective as the standard antibiotic for the treatment of suspected or confirmed MRSA?'' Because infection site, MRSA confirmation, and study year can potentially affect the choice of investigated therapies and the outcomes, these three covariates (labeled $W_i$) should be adjusted for in order to minimize confounding bias. 

The outcome of interest is clinical test of cure for all subjects who received at least one dose of treatment (a standard measure in infectious disease research). Four papers evaluated the outcome only on a subset of patients selected post-randomization; as this does not conform to our definition of the RCT-specific parameter of interest, we considered these outcomes missing. For our analysis, we chose to compare vancomycin with the two most prevalent alternatives: telavancin and linezolid. In total, 47 study arms evaluated one of these three treatments and 36 had an observed outcome. Of the remaining treatments, tigecycline, daptomycin, and ceftaroline were each evaluated in three study arms, and a regime of quinupristin/dalfopristin was evaluated in one arm. All of this information is available in the data extraction Table~\ref{extract}.

We ran four methods to obtain estimates of the counterfactual relative risk of both contrasts with the comparator vancomycin. The methods are 1) a ratio of the unadjusted mean outcomes using all available arms (called ``No Adjust''), 2) a random effects regression for the arm-specific study outcomes using a log-link and a study-specific intercept (``RE Arm''), 3) G-Computation where a random effects logistic regression weighted by the inverse standard errors is used to predict the conditional mean outcomes, and 4) TMLE with a weighted logistic random effects model for the outcome and LASSO-penalized logistic regressions (to handle the sparse data) for the propensity score and a missing data model using the R library glmnet~\citep{Friedman:001}. IPTW behaved erratically and was not included in this example. The missing outcomes required that the TMLE algorithm include fitting a model to estimate the probability of a missing outcome in each study; the TMLE update step was therefore modified to use a product of the propensity score and the probability of observing the outcome in place of $g_a(W_i)$. To estimate the standard errors and confidence intervals, the built-in functions in the library lme4 were used for the random effects model, and the clustered nonparametric bootstrap (1000 times 54 resamples of 27 studies with replacement) was used for the other methods. 

The results of the network meta-analysis are presented graphically in Figure~\ref{Figure3:sub} (and numerically in the Appendix Figure~\ref{exampleres}). We also included the results of the studies that contrasted the two treatments directly. For the comparison of telavancin versus vancomycin, all estimators include the null in the confidence interval. The random effect regression and G-Computation produce estimates of the relative risk close to one, indicating near equivalence of treatments while the point estimate of TMLE was further from the null (in the direction of the superiority of vancomycin). Notably, the confidence interval for the TMLE  in the first contrast is much wider than the others. The unadjusted method produced a point estimate in the direction of the superiority of telavancin, demonstrating that the correction for study-level confounding impacted the analysis. For the comparison of linezolid versus vancomycin, the random effects regression, G-Computation and TMLE agree on the superiority of linezolid. The original study by \citet{Bally:001} also found some suggestion of a superior effect of linezolid compared to vancomycin but for both subpopulations the confidence intervals were large and spanned the null.

\begin{figure}[ht]
\begin{center}
\subfigure[] 
{
    \label{Figure3:a}
    \includegraphics[width=0.47\textwidth]{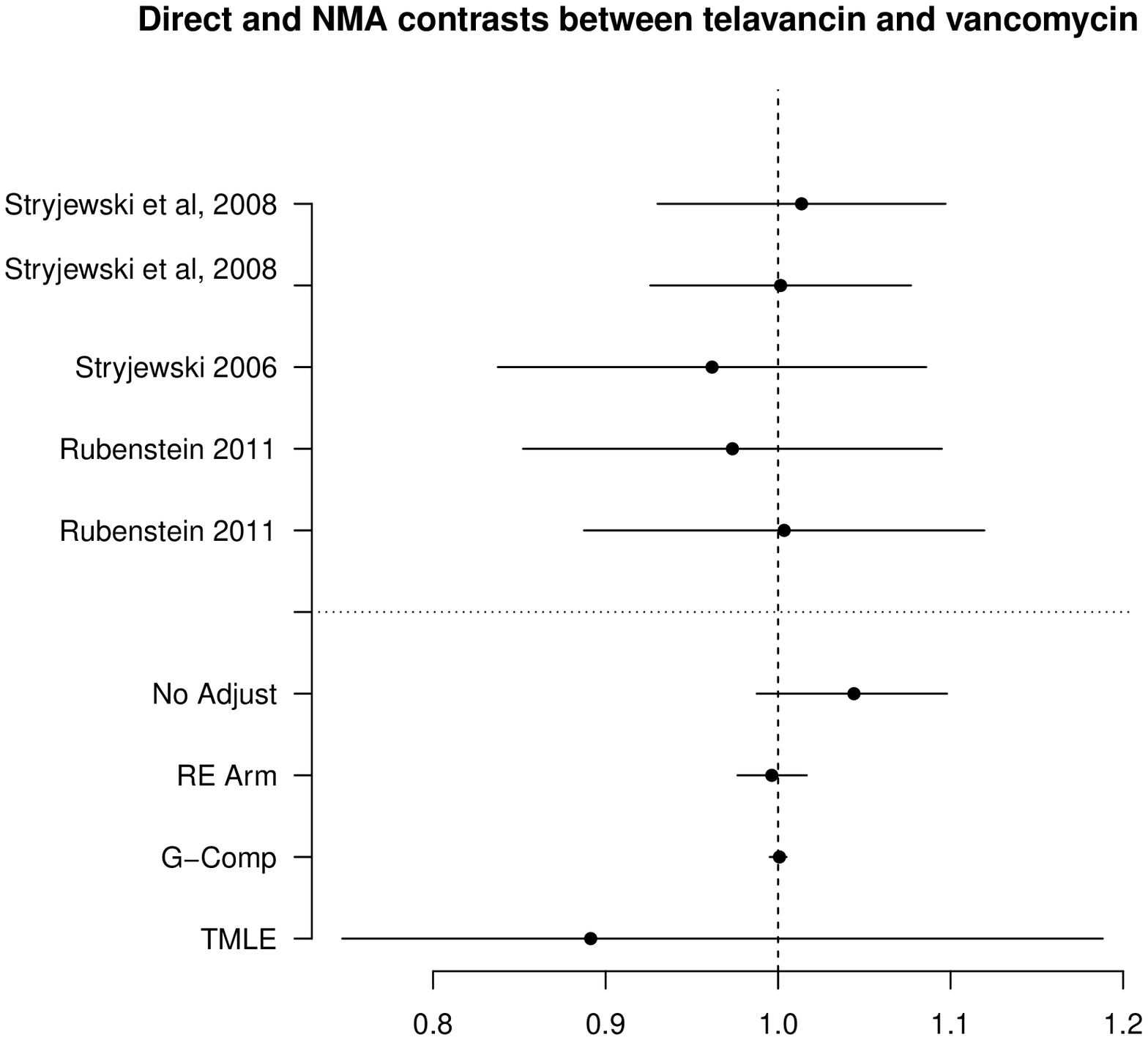}
}
\subfigure[] 
{
    \label{Figure3:b}
    \includegraphics[width=0.47\textwidth]{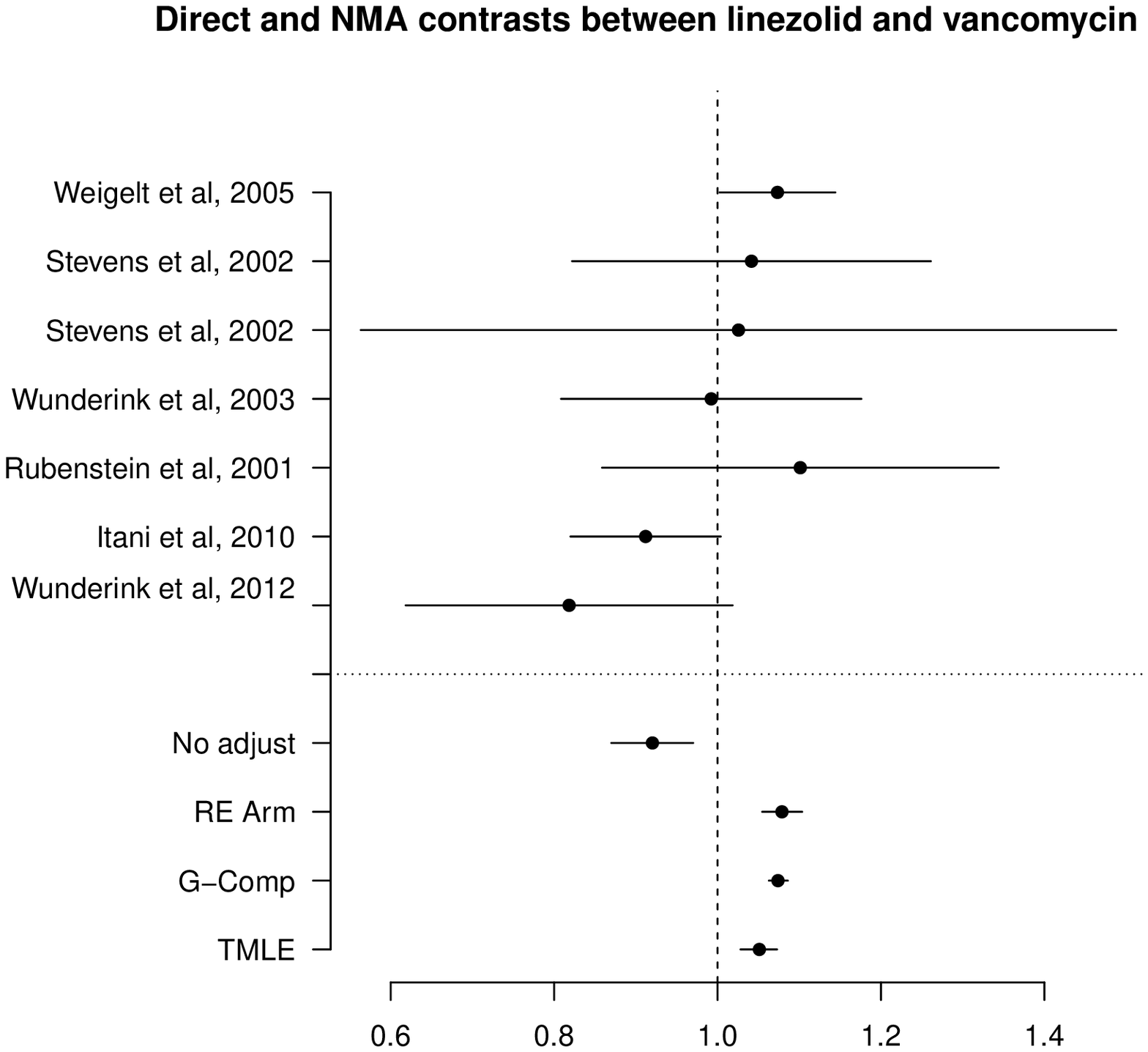}
}
\end{center}
\caption{Risk ratio estimates and confidence intervals for clinical success at test of cure for all studies with direct comparisons and all network meta-analysis methods for the contrasts between a) telavancin and vancomycin and b) linezolid and vancomycin. Risk ratio values below one indicate superiority of vancomycin.}\label{Figure3:sub}
\end{figure}

We can also easily obtain estimates of the contrast between telavancin and linezolid. The G-Computation and TMLE produce risk ratios for clinical success of 0.94 ($95\%$ confidence interval = 0.92,0.94) and 0.85 (0.71,1.12) respectively, with G-Computation concluding the superiority of linezolid. As no RCT directly contrasted these two antibiotics, this demonstrates another general advantage of network meta-analysis, which is the ability to formally compare treatments using only indirect evidence of their relative performance.

If we are to interpret the summary statistics as estimates of the relative causal effects of antibiotic choice on successful treatment, the causal assumptions in Section~\ref{assumptions} need to be satisfied. Each of the studies evaluated the clinical efficacy of the treatments, which is defined on patients who had received at least one dose of the study drug. Because randomized treatment was first-line therapy (administered intravenously in-hospital) and the success of treatment was determined clinically, each trial estimated the relative effect under full adherence.  \emph{No interference}: No interference is credible in this case because all subjects were already suspected or confirmed to have MRSA upon entry to the study. Therefore, the choice of treatment in the other arm wouldn't have an effect on existing infections nor the success of treatment. \emph{Unconfoundedness}: The unconfoundedness assumption relies on whether year, infection type, and whether MRSA was confirmed were sufficient to control for confounding at the study-level. This assumption could be violated if prognostic demographic variables were involved in the study design stage. However, prognostic markers such as diabetes and peripheral vascular disease (for cSSTI) and mechanical ventilation, APACHE II score, clinical markers of severity, and presence of organ dysfunction (for pneumonia) are unlikely to determine the choice of initial therapy~\citep{Lipsky:001,Niederman:001}. \emph{Consistency}: The dosage regimens varied somewhat across studies but were all considered to be at therapeutic levels. 
However, the length of time to the evaluation time point for each treatment type varied within and between studies (e.g. 7-14 days for telavancin versus 12-28 days for linezolid). If this corresponds to meaningfully different treatment durations (and/or periods of time lapsed before evaluation), this would indicate different definitions of interventions across studies, and thus a violation of the consistency assumption. \emph{Positivity}: All subjects in the study were indicated to receive any of the treatments evaluated.

\section{Summary}

In this paper, we nonparametrically define the parameter of interest in a network meta-analysis with direct and indirect comparisons using the counterfactual framework often employed in causal inference. This definition of the parameter of interest is model-independent and is interpretable on what we define as a metapopulation, the union of all superpopulations. Such an approach allows for a straight-forward description of what is being estimated, which is accessible even without an understanding of the estimation methods being used. In particular, we can interpret the marginal effects defined in this paper as the relative mean outcome had all subjects in the metapopulation been assigned to each treatment versus another. If a specific population is of interest and not represented by the metapopulation, with some conditions it may also be possible to more generally transport effect estimation, as described by~\citet{Bareinboim:001}.

We have presented a set of conditions under which identifiability of the parameter of interest is possible. Identifiability allows for a clear description of when the parameter of interest can and cannot be estimated. For instance, the non-interference requirement casts doubt on the synthesis of studies that allow for treatment switching, crossover, or group contamination. The assumptions that we made allowed for the simplification of the relevant components of the observed data likelihood so that arm-based inference is possible. 

One might alternatively specify the RCT-estimated contrast as the ``outcome of interest'' (rather than use the arm-specific outcome as we did). However, under this alternative, the propensity score would then be defined as the probability of a trial directly contrasting a given treatment pair. For standard network meta-analysis sample sizes, this would most often produce practical positivity problems, indicating the need for extrapolation using the outcome model (and thereby creating estimators that are very sensitive to model misspecification). In particular, two treatments that had never been directly compared would have no data support in this model.

If all treatments are selected completely at random into studies (or if only two treatments have ever been available to compare) then a standard unadjusted analysis using those arms assigned the desired treatments would be consistent. If we weaken this assumption and replace it with conditional exchangeability, then the estimators introduced in this paper are appropriate in that they allow for the adjustment of study-level covariates.  

Our methods also allow for a wider inclusion criteria of studies in a systematic review. It is often the case that systematic reviews will exclude studies because they do not evaluate the exact desired clinical endpoint. Using our proposed methods, we can avoid selection bias due to studies excluded only for this reason. To do so, we would artificially censor the outcomes of studies that do not estimate the desired outcome-type of interest. The censored outcomes of these studies might then be considered ``missing at random'' conditional on the study baseline information which should still be included in the analysis (both in the propensity score model and the missing data model).

For the analysis of continuous individual-level outcomes, we assumed independence between the sample mean and standard deviation within each study arm. While we chose to present our identifiability argument under this assumption, it is not ultimately necessary. However, it is not straight-forward to propose a valid Monte Carlo or Bayesian estimation approach to the setting with dependent sample means and standard deviations. In some cases, it may be possible to transform the individual-level data to remove the skew, but this relies on access to each study's raw data, in which case an individual patient data analysis would be preferable. 

In the simulation study, we show that certain estimators adopted from the causal inference literature can produce valid estimates of effect contrasts under the identifiability conditions described. In particular, G-Computation and TMLE might lend themselves well to network meta-analysis, which is characterized by small sample sizes and low prevalence for certain treatments. IPTW was seen to be sensitive to rare treatment assignment and G-Computation and TMLE were seen to be somewhat sensitive to model misspecification. Some general benefits of using TMLE are that it is double robust and can incorporate nonparametric (or machine learning) estimation of the propensity score and outcome model which can help avoid bias from model misspecification~\citep{vdl:006}. More methods development and investigations are needed to address extremely rare treatments and how (or whether) TMLE can be adapted to be robust in this setting.

The application we presented compared the results of random effects regression, G-Computation, and TMLE in a network meta-analysis of the relative efficacy of treatment options for MRSA infection. The random effects regression and G-Computation produced small confidence intervals relative to the direct contrasts of the individual RCTs though TMLE only did for one comparison investigated. In contrast to the analysis in the original article that used unadjusted contrast-based hierarchical Bayesian modeling on the separate subpopulations of infection types, our analyses concluded that there is evidence to support the superiority of linezolid over vancomycin. 
We also noted the poor stability of IPTW in this example and generally do not recommend this estimator when the data support for certain treatment levels is sparse. Finally, using this data example, we demonstrated how the causal assumptions should be listed and critiqued in order to stimulate discussion about the appropriateness of causal interpretations in specific contexts.


The framework we present formally assumes that we are restricting our analyses to studies evaluating a common parameter-type. If there was only partial-adherence in the RCTs, our framework does not allow for the mixing of intent-to-treat parameter estimates with adherence-adjusted parameter estimates. (Estimation of the adherence-adjusted parameters in RCTs is described in~\citealt{Hernan:002}.) 
The same restriction applies to the results of observational studies if the parameter type estimated in the observational study is not the same as in the clinical trials. Specifically, treatment adherence and outcome need to be defined identically across studies, and all studies whose endpoints are included must estimate the same mean treatment-specific counterfactual outcome. Although it is common practice to include different parameter types in a meta-analysis, our formalization of the target parameter reveals that a causal interpretation of the resulting effect estimate may be quite challenging.

 In addition to the issues we describe, there are many other concerns about aggregating study results in various settings. For instance, one might question the independence between RCTs happening close in time, or the systematic review inclusion criteria.  We believe our framework provides additional structure to the ongoing discussion about the validity of network meta-analysis and will help stimulate solutions to the remaining challenges.

\bibliographystyle{DeGruyter}
\bibliography{netmeta}

\begin{thebibliography}{17}
\newcommand{\enquote}[1]{``#1''}
\providecommand{\natexlab}[1]{#1}
\providecommand{\url}[1]{\texttt{#1}}
\providecommand{\urlprefix}{URL }

\bibitem[{Arbeit et~al.(2004)Arbeit, Maki, Tally, Campanaro, Eisenstein, and
  {Daptomycin 98-01 and 99-01 Investigators}}]{Arbeit:001}
Arbeit, R.~D., D.~Maki, F.~P. Tally, E.~Campanaro, B.~I. Eisenstein, and
  {Daptomycin 98-01 and 99-01 Investigators} (2004): \enquote{The safety and
  efficacy of daptomycin for the treatment of complicated skin and
  skin-structure infections,} \emph{Clinical Infectious Diseases}, 38,
  1673--1681.

\bibitem[{Breedt et~al.(2005)Breedt, Teras, Gardovskis, Maritz, Vaasna, Ross,
  Gioud-Paquet, Dartois, Ellis-Grosse, Loh, and {Tigecycline 305 cSSSI Study
  Group}}]{Breedt:001}
Breedt, J., J.~Teras, J.~Gardovskis, F.~J. Maritz, T.~Vaasna, D.~P. Ross,
  M.~Gioud-Paquet, N.~Dartois, E.~J. Ellis-Grosse, E.~Loh, and {Tigecycline 305
  cSSSI Study Group} (2005): \enquote{Safety and efficacy of tigecycline in
  treatment of skin and skin structure infections: results of a double-blind
  phase 3 comparison study with vancomycin-aztreonam,} \emph{Antimicrobial
  Agents and Chemotherapy}, 49, 4658--4666.

\bibitem[{Corey et~al.(2010)Corey, Wilcox, Talbot, Thye, Friedland, Baculik,
  and {CANVAS 1 investigators}}]{Corey:001}
Corey, G.~R., M.~H. Wilcox, G.~H. Talbot, D.~Thye, D.~Friedland, T.~Baculik,
  and {CANVAS 1 investigators} (2010): \enquote{Canvas 1: the first phase iii,
  randomized, double-blind study evaluating ceftaroline fosamil for the
  treatment of patients with complicated skin and skin structure infections,}
  \emph{Journal of Antimicrobial Chemotherapy}, 65 Suppl 4, iv41--51.

\bibitem[{Fagon et~al.(2000)Fagon, Patrick, Haas, Torres, Gibert, Cheadle,
  Falcone, Anholm, Paganin, Fabian, and Lilienthal}]{Fagon:001}
Fagon, J., H.~Patrick, D.~W. Haas, A.~Torres, C.~Gibert, W.~G. Cheadle, R.~E.
  Falcone, J.~D. Anholm, F.~Paganin, T.~C. Fabian, and F.~Lilienthal (2000):
  \enquote{Treatment of gram-positive nosocomial pneumonia. prospective
  randomized comparison of quinupristin/dalfopristin versus vancomycin.
  nosocomial pneumonia group,} \emph{American Journal of Respiratory and
  Critical Care Medicine}, 161, 753--762.

\bibitem[{Florescu et~al.(2008)Florescu, Beuran, Dimov, Razbadauskas, Bochan,
  Fichev, Dukart, Babinchak, Cooper, Ellis-Grosse, Dartois, Gandjini, and {307
  Study Group}}]{Florescu:001}
Florescu, I., M.~Beuran, R.~Dimov, A.~Razbadauskas, M.~Bochan, G.~Fichev,
  G.~Dukart, T.~Babinchak, C.~A. Cooper, E.~J. Ellis-Grosse, N.~Dartois,
  H.~Gandjini, and {307 Study Group} (2008): \enquote{Efficacy and safety of
  tigecycline compared with vancomycin or linezolid for treatment of serious
  infections with methicillin-resistant staphylococcus aureus or
  vancomycin-resistant enterococci: a phase 3, multicentre, double-blind,
  randomized study,} \emph{Journal of Antimicrobial Chemotherapy}, 62 Suppl 1,
  i17--28.

\bibitem[{Itani et~al.(2010)Itani, Dryden, Bhattacharyya, Kunkel, Baruch, and
  Weigelt}]{Itani:001}
Itani, K.~M., M.~S. Dryden, H.~Bhattacharyya, M.~J. Kunkel, A.~M. Baruch, and
  J.~A. Weigelt (2010): \enquote{Efficacy and safety of linezolid versus
  vancomycin for the treatment of complicated skin and soft-tissue infections
  proven to be caused by methicillin-resistant staphylococcus aureus,}
  \emph{American Journal of Surgery}, 199, 804--816.

\bibitem[{Katz et~al.(2008)Katz, Lindfield, Steenbergen, Benziger, Blackerby,
  Knapp, and Martone}]{Katz:001}
Katz, D.~E., K.~C. Lindfield, J.~N. Steenbergen, D.~P. Benziger, K.~J.
  Blackerby, A.~G. Knapp, and W.~J. Martone (2008): \enquote{A pilot study of
  high-dose short duration daptomycin for the treatment of patients with
  complicated skin and skin structure infections caused by gram-positive
  bacteria,} \emph{International Journal of Clinical Practice}, 62, 1455--1464.

\bibitem[{Rubinstein et~al.(2001)Rubinstein, Cammarata, Oliphant, Wunderink,
  and {Linezolid Nosocomial Pneumonia Study Group}}]{Rubenstein:001}
Rubinstein, E., S.~Cammarata, T.~Oliphant, R.~Wunderink, and {Linezolid
  Nosocomial Pneumonia Study Group} (2001): \enquote{Linezolid (pnu-100766)
  versus vancomycin in the treatment of hospitalized patients with nosocomial
  pneumonia: a randomized, double-blind, multicenter study,} \emph{Clinical
  Infectious Diseases}, 32, 402--412.

\bibitem[{Rubinstein et~al.(2011)Rubinstein, Lalani, Corey, Kanafani, Nannini,
  Rocha, Rahav, Niederman, Kollef, Shorr, Lee, Lentnek, Luna, Fagon, Torres,
  Kitt, Genter, Barriere, Friedland, Stryjewski, and {ATTAIN Study
  Group}}]{Rubenstein:002}
Rubinstein, E., T.~Lalani, G.~R. Corey, Z.~A. Kanafani, E.~C. Nannini, M.~G.
  Rocha, G.~Rahav, M.~S. Niederman, M.~H. Kollef, A.~F. Shorr, P.~C. Lee, A.~L.
  Lentnek, C.~M. Luna, J.~Y. Fagon, A.~Torres, M.~M. Kitt, F.~C. Genter, S.~L.
  Barriere, H.~D. Friedland, M.~E. Stryjewski, and {ATTAIN Study Group} (2011):
  \enquote{Telavancin versus vancomycin for hospital-acquired pneumonia due to
  gram-positive pathogens,} \emph{Clinical Infectious Diseases}, 52, 31--40.

\bibitem[{Sacchidanand et~al.(2005)Sacchidanand, Penn, Embil, Campos, Curcio,
  Ellis-Grosse, Loh, and Rose}]{Sacchidanand:001}
Sacchidanand, S., R.~L. Penn, J.~M. Embil, M.~E. Campos, D.~Curcio,
  E.~Ellis-Grosse, E.~Loh, and G.~Rose (2005): \enquote{Efficacy and safety of
  tigecycline monotherapy compared with vancomycin plus aztreonam in patients
  with complicated skin and skin structure infections: Results from a phase 3,
  randomized, double-blind trial,} \emph{International Journal of Infectious
  Diseases}, 9, 251--261.

\bibitem[{Stryjewski et~al.(2006)Stryjewski, Chu, O'Riordan, Warren, Dunbar,
  Young, Vall\'ee, Fowler, Morganroth, Barriere, Kitt, Corey, and {FAST 2
  Investigator Group}}]{Stryjewski:002}
Stryjewski, M.~E., V.~H. Chu, W.~D. O'Riordan, B.~L. Warren, L.~M. Dunbar,
  D.~M. Young, M.~Vall\'ee, V.~G.~J. Fowler, J.~Morganroth, S.~Barriere, M.~M.
  Kitt, G.~R. Corey, and {FAST 2 Investigator Group} (2006):
  \enquote{Telavancin versus standard therapy for treatment of complicated skin
  and skin structure infections caused by gram-positive bacteria: Fast 2
  study,} \emph{Antimicrobial Agents and Chemotherapy}, 50, 862--867.

\bibitem[{Stryjewski et~al.(2008)Stryjewski, Graham, Wilson, O'Riordan, Young,
  Lentnek, Ross, Fowler, Hopkins, Friedland, Barriere, Kitt, Corey, and
  {Assessment of Telavancin in Complicated Skin and Skin-Structure Infections
  Study}}]{Stryjewski:001}
Stryjewski, M.~E., D.~R. Graham, S.~E. Wilson, W.~O'Riordan, D.~Young,
  A.~Lentnek, D.~P. Ross, V.~G. Fowler, A.~Hopkins, H.~D. Friedland, S.~L.
  Barriere, M.~M. Kitt, G.~R. Corey, and {Assessment of Telavancin in
  Complicated Skin and Skin-Structure Infections Study} (2008):
  \enquote{Telavancin versus vancomycin for the treatment of complicated skin
  and skin-structure infections caused by gram-positive organisms,}
  \emph{Clinical Infectious Diseases}, 46, 1683--1693.

\bibitem[{Talbot et~al.(2007)Talbot, Thye, Das, and Ge}]{Talbot:001}
Talbot, G.~H., D.~Thye, A.~Das, and Y.~Ge (2007): \enquote{Phase 2 study of
  ceftaroline versus standard therapy in treatment of complicated skin and skin
  structure infections,} \emph{Antimicrobial Agents and Chemotherapy}, 51,
  3612--3616.

\bibitem[{Wilcox et~al.(2010)Wilcox, Corey, Talbot, Thye, Friedland, Baculik,
  and {CANVAS 2 investigators}}]{Wilcox:001}
Wilcox, M.~H., G.~R. Corey, G.~H. Talbot, D.~Thye, D.~Friedland, T.~Baculik,
  and {CANVAS 2 investigators} (2010): \enquote{Canvas 2: the second phase iii,
  randomized, double-blind study evaluating ceftaroline fosamil for the
  treatment of patients with complicated skin and skin structure infections,}
  \emph{Journal of Antimicrobial Chemotherapy}, 65 Suppl 4, iv53--65.

\bibitem[{Wunderink et~al.(2003)Wunderink, Cammarata, Oliphant, Kollef, and
  {Linezolid Nosocomial Pneumonia Study Group}}]{Wunderink:002}
Wunderink, R.~G., S.~K. Cammarata, T.~H. Oliphant, M.~H. Kollef, and {Linezolid
  Nosocomial Pneumonia Study Group} (2003): \enquote{Continuation of a
  randomized, double-blind, multicenter study of linezolid versus vancomycin in
  the treatment of patients with nosocomial pneumonia,} \emph{Clinical
  Therapeutics}, 25, 980--992.

\bibitem[{Wunderink et~al.(2008)Wunderink, Mendelson, Somero, Fabian, May,
  Bhattacharyya, Leeper, and Solomkin}]{Wunderink:001}
Wunderink, R.~G., M.~H. Mendelson, M.~S. Somero, T.~C. Fabian, A.~K. May,
  H.~Bhattacharyya, K.~V.~J. Leeper, and J.~S. Solomkin (2008): \enquote{Early
  microbiological response to linezolid vs vancomycin in ventilator-associated
  pneumonia due to methicillin-resistant staphylococcus aureus,} \emph{Chest},
  134, 1200--1207.

\bibitem[{Wunderink et~al.(2012)Wunderink, Niederman, Kollef, Shorr, Kunkel,
  Baruch, McGee, Reisman, and Chastre}]{Wunderink:003}
Wunderink, R.~G., M.~S. Niederman, M.~H. Kollef, A.~F. Shorr, M.~J. Kunkel,
  A.~Baruch, W.~T. McGee, A.~Reisman, and J.~Chastre (2012): \enquote{Linezolid
  in methicillin-resistant staphylococcus aureus nosocomial pneumonia: a
  randomized, controlled study,} \emph{Clinical Infectious Disease}, 54,
  621--629.

\end{thebibliography}


\begin{thebibliography}{44}
\newcommand{\enquote}[1]{``#1''}
\providecommand{\natexlab}[1]{#1}
\providecommand{\url}[1]{\texttt{#1}}
\providecommand{\urlprefix}{URL }

\bibitem[{Alonso et~al.(2015)Alonso, Van~der Elst, Molenberghs, Buyse, and
  Burzykowski}]{Alonso:001}
Alonso, A., W.~Van~der Elst, G.~Molenberghs, M.~Buyse, and T.~Burzykowski
  (2015): \enquote{On the relationship between the causal-inference and
  meta-analytic paradigms for the validation of surrogate endpoints,}
  \emph{Biometrics}, 71, 15--24.

\bibitem[{Bally et~al.(2012)Bally, Dendukuri, Sinclair, Ahern, Poisson, and
  Brophy}]{Bally:001}
Bally, M., N.~Dendukuri, A.~Sinclair, S.~P. Ahern, M.~Poisson, and J.~Brophy
  (2012): \enquote{A network meta-analysis of antibiotics for treatment of
  hospitalised patients with suspected or proven meticillin-resistant
  \emph{Staphylococcus aureus} infection,} \emph{International Journal of
  Antimicrobial Agents}, 40, 479--495.

\bibitem[{Bareinboim and Pearl(2013)}]{Bareinboim:001}
Bareinboim, E. and J.~Pearl (2013): \enquote{Meta-transportability of causal
  effects: {A} formal approach,} in \emph{Proceedings of the 16th
  {I}nternational {C}onference on {A}rtificial {I}ntelligence and
  {S}tatistics}.

\bibitem[{Berlin and Golub(2014)}]{Berlin:001}
Berlin, J.~A. and R.~M. Golub (2014): \enquote{Meta-analysis as evidence:
  {B}uilding a better pyramid,} \emph{Journal of the American Medical
  Association}, 312, 603--606.

\bibitem[{Caldwell et~al.(2005)Caldwell, Ades, and Higgins}]{Caldwell:001}
Caldwell, D.~M., A.~E. Ades, and J.~P.~T. Higgins (2005): \enquote{Simultaneous
  comparison of multiple treatments: combining direct and indirect evidence,}
  \emph{BMJ}, 331, 897--900.

\bibitem[{Cole and Frangakis(2009)}]{Cole:003}
Cole, S.~R. and C.~E. Frangakis (2009): \enquote{The consistency statement in
  causal inference: A definition or an assumption?} \emph{Epidemiology}, 20,
  3--5.

\bibitem[{Cole and Hern\'{a}n(2008)}]{Cole:002}
Cole, S.~R. and M.~A. Hern\'{a}n (2008): \enquote{Constructing inverse
  probability weights for marginal structural models,} \emph{American Journal
  of Epidemiology}, 168, 656--664.

\bibitem[{Cope et~al.(2014)Cope, Zhang, Saletan, Smiechowski, Jansen, and
  Schmid}]{Cope:001}
Cope, S., J.~Zhang, S.~Saletan, B.~Smiechowski, J.~P. Jansen, and P.~Schmid
  (2014): \enquote{A process for assessing the feasibility of a network
  meta-analysis: a case study of everolimus in combination with hormonal
  therapy versus chemotherapy for advanced breast cancer,} \emph{BMC Medicine},
  12.

\bibitem[{Dias et~al.(2013{\natexlab{a}})Dias, Sutton, Ades, and
  Welton}]{Dias:001}
Dias, S., A.~J. Sutton, A.~E. Ades, and N.~J. Welton (2013{\natexlab{a}}):
  \enquote{A generalized linear modeling framework for pairwise and network
  meta-analysis of randomized controlled trials,} \emph{Medical Decision
  Making}, 33, 607--617.

\bibitem[{Dias et~al.(2013{\natexlab{b}})Dias, Sutton, Welton, and
  Ades}]{Dias:003}
Dias, S., A.~J. Sutton, N.~J. Welton, and A.~E. Ades (2013{\natexlab{b}}):
  \enquote{Evidence synthesis for decision making 3: Heterogeneity—subgroups,
  meta-regression, bias, and bias-adjustment,} \emph{Medical Decision Making},
  33, 618--640.

\bibitem[{Efron and Tibshirani(1994)}]{Efron:001}
Efron, B. and R.~J. Tibshirani (1994): \emph{An Introduction to the Bootstrap},
  Monographs on Statistics and Applied Probability, Chapman \& Hall/CRC.

\bibitem[{Ferguson(1996)}]{Ferguson}
Ferguson, T.~S. (1996): \emph{A Course in Large Sample Theory}, Texts in
  Statistical Science, Chapman \& Hall/CRC.

\bibitem[{Friedman et~al.(2010)Friedman, Hastie, and Tibshirani}]{Friedman:001}
Friedman, J., T.~Hastie, and R.~Tibshirani (2010): \enquote{Regularization
  paths for generalized linear models via coordinate descent,} \emph{Journal of
  Statistical Software}, 33, 1--22,
  \urlprefix\url{http://www.jstatsoft.org/v33/i01/}.

\bibitem[{Gail et~al.(1984)Gail, Wieand, and Piantadosi}]{Gail:001}
Gail, M.~H., S.~Wieand, and S.~Piantadosi (1984): \enquote{Biased estimates of
  treatment effect in randomized experiments with nonlinear regressions and
  omitted covariates,} \emph{Biometrika}, 71, 431--444.

\bibitem[{Gruber and van~der Laan(2010)}]{Gruber:001}
Gruber, S. and M.~J. van~der Laan (2010): \enquote{A targeted maximum
  likelihood estimator of a causal effect on a bounded continuous outcome,}
  \emph{The International Journal of Biostatistics}, 6, Article 26.

\bibitem[{Hern\'{a}n and Hern\'andez-D\'iaz(2012)}]{Hernan:002}
Hern\'{a}n, M.~A. and S.~Hern\'andez-D\'iaz (2012): \enquote{Beyond the
  intention to treat in comparative effectiveness research,} \emph{Clinical
  trials}, 9, 48--55.

\bibitem[{Hong et~al.(2016)Hong, Chu, Zhang, and Carlin}]{Hong:001}
Hong, H., H.~Chu, J.~Zhang, and B.~P. Carlin (2016): \enquote{Rejoinder to the
  discussion of ``{A} {B}ayesian missing data framework for generalized
  multiple outcome mixed treatment comparisons,'' by {S}. {D}ias and {A.E.}
  {A}des,} \emph{Research Synthesis Methods}, 7, 29--33.

\bibitem[{Imbens(2000)}]{Imbens:001}
Imbens, G.~W. (2000): \enquote{The role of the propensity score in estimating
  dose-response functions,} \emph{Biometrika}, 87, 706--710.

\bibitem[{Jansen et~al.(2012)Jansen, Schmid, and Salanti}]{Jansen:001}
Jansen, J.~P., C.~H. Schmid, and G.~Salanti (2012): \enquote{Directed acyclic
  graphs can help understand bias in indirect and mixed treatment comparisons,}
  \emph{Journal of Clinical Epidemiology}, 65, 798--807.

\bibitem[{Jansen et~al.(2014)Jansen, Trikalinos, Cappelleri, Daw, Andes,
  Eldessouki, and Salanti}]{Jansen:002}
Jansen, P.~J., T.~Trikalinos, J.~C. Cappelleri, J.~Daw, S.~Andes,
  R.~Eldessouki, and G.~Salanti (2014): \enquote{Indirect treatment
  comparison/network meta-analysis study questionnaire to assess relevance and
  credibility to inform health care decision making: An ispor-amcp-npc good
  practice task force report,} \emph{Value in Health}, 17, 157--173.

\bibitem[{Lipsky et~al.(2011)Lipsky, Itani, Weigelt, Joseph, Paap, Reisman,
  Myers, and Huang}]{Lipsky:001}
Lipsky, B.~A., K.~M. Itani, J.~A. Weigelt, W.~Joseph, C.~M. Paap, A.~Reisman,
  D.~E. Myers, and D.~B. Huang (2011): \enquote{The role of diabetes mellitus
  in the treatment of skin and skin structure infections caused by
  methicillin-resistant staphylococcus aureus: results from three randomized
  controlled trials,} \emph{International Journal of Infectious Diseases}, 15,
  e140--e146.

\bibitem[{Liu et~al.(2011)Liu, Bayer, Cosgrove, Daum, Fridkin, Gorwitz, Kaplan,
  Karchmer, Levine, Murray, Rybak, Talan, and Chambers}]{Liu:001}
Liu, C., A.~Bayer, S.~E. Cosgrove, R.~S. Daum, S.~K. Fridkin, R.~J. Gorwitz,
  S.~L. Kaplan, A.~W. Karchmer, D.~P. Levine, B.~E. Murray, M.~J. Rybak, D.~A.
  Talan, and H.~F. Chambers (2011): \enquote{Clinical practice guidelines by
  the infectious diseases society of america for the treatment of
  methicillin-resistant staphylococcus aureus infections in adults and
  children,} \emph{Clinical Infectious Diseases}, 52, e18--e55.

\bibitem[{Lu and Ades(2004)}]{Lu:001}
Lu, G. and A.~E. Ades (2004): \enquote{Combination of direct and indirect
  evidence in mixed treatment comparisons,} \emph{Statistics in Medicine}, 23,
  3105--3124.

\bibitem[{Lu and Ades(2006)}]{Lu:002}
Lu, G. and A.~E. Ades (2006): \enquote{Assessing evidence inconsistency in
  mixed treatment comparisons,} \emph{Journal of the American Statistical
  Association}, 101, 447--459.

\bibitem[{Lumley(2004)}]{Lumley:001}
Lumley, T. (2004): \enquote{Network meta-analysis for indirect treatment
  comparisons,} \emph{Statistics in Medicine}, 21, 2313--2324.

\bibitem[{Niederman(2010)}]{Niederman:001}
Niederman, M.~S. (2010): \enquote{Hospital-acquired pneumonia, health
  care-associated pneumonia, ventilator-associated pneumonia, and
  ventilator-associated tracheobronchitis: definitions and challenges in trial
  design,} \emph{Clinical Infectious Diseases}, 51 Suppl 1, S12--S7.

\bibitem[{Pearl(2009)}]{Pearl:002}
Pearl, J. (2009): \emph{Causality}, Cambridge University Press, 2 edition.

\bibitem[{Porter et~al.(2011)Porter, Gruber, van~der Laan, and
  Sekhon}]{Porter:001}
Porter, K.~E., S.~Gruber, M.~J. van~der Laan, and J.~S. Sekhon (2011):
  \enquote{The relative performance of targeted maximum likelihood estimators,}
  \emph{The International Journal of Biostatistics}, 7, 1--34.

\bibitem[{Robins(1986)}]{Robins:002}
Robins, J.~M. (1986): \enquote{A new approach to causal inference in mortality
  studies with a sustained exposure period -- application to control of the
  healthy worker survivor effect,} \emph{Mathematical Modelling}, 7,
  1393--1512.

\bibitem[{Robins(1988)}]{Robins:009}
Robins, J.~M. (1988): \enquote{Confidence intervals for causal parameters,}
  \emph{Statistics in Medicine}, 7, 773--785.

\bibitem[{Rubin(1980)}]{Rubin:003}
Rubin, D.~B. (1980): \enquote{Randomization analysis of experimental data: The
  fisher randomization test comment,} \emph{Journal of the American Statistical
  Association}, 75, 591--593.

\bibitem[{Salanti et~al.(2008)Salanti, Higgins, Ades, and
  Ioannidis}]{Salanti:001}
Salanti, G., J.~P.~T. Higgins, A.~E. Ades, and J.~P.~A. Ioannidis (2008):
  \enquote{Evaluation of networks of randomized trials,} \emph{Statistical
  Methods in Medical Research}, 17, 279--301.

\bibitem[{Salanti et~al.(2009)Salanti, Marinho, and Higgins}]{Salanti:002}
Salanti, G., V.~Marinho, and J.~P.~T. Higgins (2009): \enquote{A case study of
  multiple-treatments meta-analysis demonstrates that covariates should be
  considered,} \emph{Journal of Clinical Epidemiology}, 62, 857--864.

\bibitem[{Schnitzer et~al.(2013)Schnitzer, Moodie, and Platt}]{Schnitzer:001}
Schnitzer, M.~E., E.~E.~M. Moodie, and R.~W. Platt (2013): \enquote{Targeted
  maximum likelihood estimation for marginal time-dependent treatment effects
  under density misspecification,} \emph{Biostatistics}, 14, 1--14.

\bibitem[{Slavin(1995)}]{Slavin:001}
Slavin, R.~E. (1995): \enquote{Best evidence synthesis: An intelligent
  alternative to meta-analysis,} \emph{Journal of Clinical Epidemiology}, 48,
  9--18.

\bibitem[{Snowden et~al.(2011)Snowden, Rose, and Mortimer}]{Snowden:001}
Snowden, J.~M., S.~Rose, and K.~M. Mortimer (2011): \enquote{Implementation of
  g-computation on a simulated data set: Demonstration of a causal inference
  technique,} \emph{American Journal of Epidemiology}, 173, 731--738.

\bibitem[{Tsiatis(2006)}]{Tsiatis}
Tsiatis, A.~A. (2006): \emph{Semiparametric Theory and Missing Data}, Springer
  Series in Statistics, Springer.

\bibitem[{van~der Laan and Robins(2003)}]{vdl:005}
van~der Laan, M.~J. and J.~M. Robins (2003): \emph{Unified Methods for Censored
  Longitudinal Data and Causality}, Springer Series in Statistics, Springer
  Verlag: New York.

\bibitem[{van~der Laan and Rose(2011)}]{vdl:006}
van~der Laan, M.~J. and S.~Rose (2011): \emph{Targeted Learning: Causal
  Inference for Observational and Experimental Data}, Springer Series in
  Statistics, Springer.

\bibitem[{van~der Laan and Rubin(2006)}]{vdl:001}
van~der Laan, M.~J. and D.~Rubin (2006): \enquote{Targeted maximum likelihood
  learning,} \emph{The International Journal of Biostatistics}, 2, Article 11.

\bibitem[{VanderWeele and Hern\'an(2013)}]{VanderWeele:002}
VanderWeele, T.~J. and M.~A. Hern\'an (2013): \enquote{Causal inference under
  multiple versions of treatment,} \emph{Journal of Causal Inference}, 1,
  1--20.

\bibitem[{Welton et~al.(2015)Welton, Soares, Palmer, Ades, Harrison,
  Shankar-Hari, and Rowan}]{Welton:001}
Welton, N.~J., M.~O. Soares, S.~Palmer, A.~E. Ades, D.~Harrison,
  M.~Shankar-Hari, and K.~M. Rowan (2015): \enquote{Accounting for
  heterogeneity in relative treatment effects for use in cost-effectiveness
  models and value-of-information analyses,} \emph{Medical Decision Making},
  35, 608--621.

\bibitem[{Zhang et~al.(2014)Zhang, Carlin, Neaton, Soon, Nie, Kane, Virnig, and
  Chu}]{Zhang:001}
Zhang, J., B.~P. Carlin, J.~D. Neaton, G.~G. Soon, L.~Nie, R.~Kane, B.~A.
  Virnig, and H.~Chu (2014): \enquote{Network meta-analysis of randomized
  clinical trials: Reporting the proper summaries,} \emph{Clinical Trials}, 11,
  246--262.

\bibitem[{Zhang et~al.(2015)Zhang, Chu, Hong, Virnig, and Carlin}]{Zhang:002}
Zhang, J., H.~Chu, H.~Hong, B.~A. Virnig, and B.~P. Carlin (2015):
  \enquote{Bayesian hierarchical models for network meta-analysis incorporating
  nonignorable missingness,} \emph{Statistical Methods in Medical Research}.

\end{thebibliography}

\appendix
\section{Appendix}

\subsection{Proof of identifiability under structural independence}\label{App1}

The joint counterfactual distribution  can be decomposed as $f(O^a_{i})=f_1(O^a_{i})f_2(O^a_{i})$ where $f_1(O^a_i)= Q_W(W_i)Q_{\bar{Y}}(\bar{Y}_{ij}^a\mid N^a_{ij},W_{i})$ and 
\begin{align*}
f_2(O^a_{i})=& Q_{n}(n_i\mid W_i)g_{A\backslash j}(A^a_{i\backslash j}\mid n_i,W_i)Q_{N}(N_{ij}^a\mid A^a_{i\backslash j},n_i,W_i)Q_{S}(S^a_{ij}\mid N^a_{ij},W_{i})\times\\
&\prod_{j^*\neq j} Q_{\bar{Y},S}(\bar{Y}^a_{ij^*},S^a_{ij^*}\mid N^a_{ij^*},A^a_{ij^*},W_{i})Q_N(N_{ij^*}^a\mid A^a_{i\backslash j},n_i,W_i).
\end{align*}

Let $\mathcal{A}$ be the set of possible treatments. The target of our analysis is the study arm counterfactual outcome under treatment $a$, or $E(\bar{Y}^a_{ij})=M^a$.  This mean can be written as
\begin{align*}&\int_{W_i}\sum_{n_{i}=1}^{\infty} \sum_{A_{i\backslash j }\in \{\mathcal{A}\backslash a\}}\sum_{N_{i\backslash j}=1}^{\infty}\sum_{N_{ij}=1}^{\infty}\int_{0}^{\infty}\int_{-\infty}^{\infty}\int_{0}^{\infty}\int_{-\infty}^{\infty}\bar{Y}^a_{ij}f(O^a_{i})d\bar{Y}^a_{ij}dS^a_{ij}d\bar{Y}^a_{ij^*}dS^a_{ij^*}d{W_i}\\
&=\int_{W_i}\sum_{n_{i}=1}^{\infty} \sum_{A_{i\backslash j }\in \{\mathcal{A}\backslash a\}}\sum_{N_{i\backslash j}=1}^{\infty}\sum_{N_{ij}=1}^{\infty}\int_{0}^{\infty}\int_{-\infty}^{\infty}\int_{0}^{\infty}\int_{-\infty}^{\infty}\bar{Y}^a_{ij}Q_{\bar{Y}}(\bar{Y}_{ij}^a\mid N^a_{ij},W_{i})d\bar{Y}^a_{ij}\times\\
&\quad\quad\quad\quad\quad\quad\quad\quad\quad\quad\quad\quad\quad\quad\quad\quad\quad\quad\quad\quad\quad f_2(O^a_{i}) Q_W(W_i)dS^a_{ij}d\bar{Y}^a_{ij^*}dS^a_{ij^*}dW_i\\
&=\int_{W_i}\sum_{n_{i}=1}^{\infty} \sum_{A_{i\backslash j }\in \{\mathcal{A}\backslash a\}}\sum_{N_{i\backslash j}=1}^{\infty}\sum_{N_{ij}=1}^{\infty}\int_{0}^{\infty}\int_{-\infty}^{\infty}\int_{0}^{\infty}E(\bar{Y}^a_{ij}\mid N^a_{ij},W_i)f_2(O^a_{i}) Q_W(W_i)dS^a_{ij}d\bar{Y}^a_{ij^*}dS^a_{ij^*}dW_i
\end{align*}
where the integral for $W_i$ can be a multiple integral, taken over the domain of potentially multivariate $W_i$. Now we note that for identically distributed and conditionally independent draws $Y_{ijk}$
\begin{align*}E(\bar{Y}^a_{ij}\mid N^a_{ij},W_i)&=E(\frac{1}{N^a_{ij}}\sum_{k=1}^{N^a_{ij}} Y^a_{ijk}\mid  N^a_{ij},W_i)=E(Y^a_{ijk}\mid W_i)
\end{align*}
because we assume that the study size has no effect on the individual-level outcome. It follows that $E(\bar{Y}^a_{ij}\mid N^a_{ij},W_i)$ is conditionally independent of $N^a_{ij}$. The expression for $M^a$ then simplifies to $\int_{W_i} E(\bar{Y}^a_{ij}\mid N^a_{ij},W_i)Q_W(W_i)dW_i$. In order for the conditional expectation to be estimable from the observed data, we require the unconfoundedness assumption $\bar{Y}^a_{ij}\cip A_{ij}=a\mid N^a_{ij},W_i$. With respect to the example DAG in Figure~\ref{Figure2:a}, this corresponds to having measured all components of $W2_i$. If this assumption holds in addition to the consistency of treatment for $\bar{Y}_{ij}$, we may write $M^a=\int_W E(\bar{Y}_{ij}\mid W_i,A_{ij}=a)Q_W(W_i)dW$ to establish identifiability.

\subsection{Identifiability without assuming structural independence}\label{App2}
It may not be plausible to assume conditional independence between $\bar{Y}^a_{ij}$ and $S^a_{ij}$. 
In this case, the relevant part of the distribution of the observed data counterfactuals is
\begin{align*}
f_3(O^a_{i})=Q_W(W_i)Q_{n}(n_i\mid W_i)g_{A\backslash j}(A^a_{i\backslash j}\mid n_i,W_i)Q_{N}(N_{ij}^a\mid A^a_{i\backslash j},n_i,W_i)Q_{\bar{Y},S}(\bar{Y}^a_{ij},S^a_{ij}\mid N^a_{ij},W_{i}).
\end{align*}
The target parameter can be estimated as a multiple integral over each $\bar{Y}_{ij}^a$ and each density component in $f_3(O^a_{i})$. Identifiability in this case requires a list of unconfoundedness assumptions: $A^a_{i\backslash j}\cip A_{ij}=a\mid n_i,W_i$, $N_{ij}^a\mid A_{ij}=a\mid  A^a_{i\backslash j},n_i,W_i$, and $\bar{Y}^a_{ij},S^a_{ij} \cip A_{ij}=a \mid N^a_{ij},W_{i}$. Assuming the DAG in Figure~\ref{Figure2:a} in the main manuscript, this requires having measured all components of $W1_i$, $W2_i$, and $W3_i$. It also requires the consistency assumption for $A_{i\backslash j}$, $N_{ij}$, $\bar{Y}_{ij}$ and $S_{ij}$. Under these assumptions, we can rewrite the relevant density component as 
\begin{align*}
f_3(O^a_{i})=&Q_W(W_i)Q_{n}(n_i\mid W_i)g_{A\backslash j}(A_{i\backslash j}\mid A_{ij}=a,n_i,W_i)Q_{N}(N_{ij}\mid A_{ij}=a,A_{i\backslash j},n_i,W_i)\times\\
&Q_{\bar{Y},S}(\bar{Y}_{ij},S_{ij}\mid A_{ij}=a,N_{ij},W_{i}).
\end{align*}
Since each component of this density is estimable from the data, we have identifiability of the target parameter in this case as well.

\subsection{Efficiency and Consistency of TMLE}\label{AppTMLE}

The local semiparametric efficiency and estimation consistency of the TMLE we describe can be derived very similarly to the standard observational data setting (with a single categorical exposure variable) for the estimation of the average treatment effect~\citep{vdl:006}. To give more insight into how this extends to the network meta-analysis case, we present some additional details and a proof of double robustness. 

The efficient influence function for parameter of interest $M^a$ with only aggregate observed data is
\[D_{ij}^*(O)= \left\{\bar{Y}_{ij}-E(\bar{Y}_i^a\mid W_i, a\in A_i)\right\}\frac{\mathbb{I}(a \in A_i)}{g_a(W_i)} + E(\bar{Y}_i^a\mid W_i, a\in A_i) - M^a.\]
Note that the TMLE update step produces values of $\hat{\bar{Y}}^{a,*}_{ij}$ that solve the empirical efficient influence function equation:
\begin{align*}
&\sum_{i=1}^{N}\sum_{j=1}^{n_i} (\bar{Y}_{ij}-\hat{\bar{Y}}^{a,*}_{ij})\frac{\mathbb{I}(a \in A_i)}{g_a(W_i)} + (\hat{\bar{Y}}^{a,*}_{ij} - \hat{M}_{TMLE}^a)=0
\end{align*}
so that it follows that the TMLE is a locally efficient estimator~\citep{vdl:005, Tsiatis}.
Specifically, the logistic regression update step with single covariate $X_i=\mathbb{I}(a \in A_i)/g_a(W_i)$ solves the score equation $\sum_{i=1}^{N}\sum_{j=1}^{n_i} X_i(\bar{Y}_{ij}-\hat{\bar{Y}}^{a,*}_{ij})=0$ and in the last TMLE step we set $\hat{M}_{TMLE}^a=\sum_{i=1}^{N}\sum_{j=1}^{n_i} \hat{\bar{Y}}^{a,*}_{ij}$.

First suppose that for increasing values of $\sum_{i=1}^N \mathbb{I}(a \in A_i)$, the generalized propensity score $g_a(W_i)$ converges to some $\tilde{g}_a(W_i) \neq P(a\in A_i\mid W_i)$ but that $\hat{\bar{Y}}^{a,*}_{ij}$ converges to the true values $E(\bar{Y}_{ij} \mid W_i, a\in A_i)$. We then have that 
\begin{align*}
&E\left[\left\{\bar{Y}_{ij}-E(\bar{Y}_i^a\mid W_i, a\in A_i)\right\}\times \frac{\mathbb{I}(a \in A_i)}{\tilde{g}_a(W_i)} + E(\bar{Y}_i^a\mid W_i, a\in A_i) - M^a \right]\\
&=E\left[ E\left\{ \bar{Y}_{ij}-E(\bar{Y}_i^a\mid W_i, a\in A_i) \mid W_i, a\in A_i \right\}\times \frac{\mathbb{I}(a \in A_i)}{\tilde{g}_a(W_i)}+ E(\bar{Y}_i^a\mid W_i, a\in A_i) - M^a \right]\\
&=E\left[ 0\times \frac{\mathbb{I}(a \in A_i)}{\tilde{g}_a(W_i)}\right]+ 0 = 0
\end{align*}

Now suppose that $g_a(W_i)$ converges to the true values $P(a\in A_i\mid W_i)$ but that $\hat{\bar{Y}}^{a,*}_{ij}$ converges to some function $\tilde{Q}_a(W_i)\neq E(\mid{Y}_{ij} \mid W_i, a\in A_i)$. We then have that 
\begin{align*}
&E\left[\left\{\bar{Y}_{ij}-\tilde{Q}_a(W_i)\right\}\times \frac{\mathbb{I}(a \in A_i)}{P(a\in A_i\mid W_i)} + \tilde{Q}_a(W_i) - M^a \right]\\
&=E\left[ \left\{\bar{Y}_{ij}-\tilde{Q}_a(W_i)\right\} \times E\left\{ \frac{\mathbb{I}(a \in A_i)}{P(a\in A_i\mid W_i)} \right\} + \tilde{Q}_a(W_i) - M^a \right]\\
&=E\left[ \left\{\bar{Y}_{ij}-\tilde{Q}_a(W_i)\right\} \times 1 + \tilde{Q}_a(W_i) - M^a \right]\\
&= E\left[ \bar{Y}_{ij} - M^a \right] = 0
\end{align*}
Therefore, if either of the models for $E(\bar{Y}_i^a\mid W_i, a\in A_i)$ or $P(a\in A_i\mid W_i)$ are consistent, then the TMLE for $M^a$ is also consistent as the efficient influence function equation is consistent for $M^a$.

\subsection{Data extraction information and numerical results for the example of antibiotic use on methicillin-resistant \emph{Staphylococcus aureus} infection}

Table~\ref{extract} presents the full study list from the systematic review of~\citet{Bally:001} and the data that we used in the analysis in Section~\ref{application}. Table~\ref{exampleres} presents the numerical results that we obtained from our analyses, corresponding with Figure~\ref{Figure3:sub}. The full reference list is below.

\begin{table}[]\scriptsize{
\centering
\caption{Data extraction table for the network meta-analysis of antibiotic use on methicillin-resistant \emph{Staphylococcus aureus} infection}
\label{extract}
\begin{tabular}{lccccccc}
\hline
Publication                      & Events & Ni  & Ai                                                                   & StudyID & Year & Infection & Confirmed MRSA \\
&&&&&&&at baseline\\
\hline
Katz et al., 2008      & 42     & 48  & vancomycin                                                           & 1       & 2007 & cSSTI     & 0              \\
    & 36     & 48  & daptomycin                                                           & 1       & 2007 & cSSTI     & 0              \\
Arbeit et al., 2004       & 162    & 266 & vancomycin                                                           & 2       & 2001 & cSSTI     & 0              \\
    & 165    & 264 & daptomycin                                                           & 2       & 2001 & cSSTI     & 0              \\
   & 235    & 292 & vancomycin                                                           & 3       & 2000 & cSSTI     & 0              \\
 & 217    & 270 & daptomycin                                                           & 3       & 2000 & cSSTI     & 0              \\
Breedtet al.,2005         & 216    & 250 & vancomycin                                                           & 4       & 2003 & cSSTI     & 0              \\
     & 212    & 253 & tigecycline                                                          & 4       & 2003 & cSSTI     & 0              \\
Sacchidanand et al., 2005 & 196    & 255 & vancomycin                                                           & 5       & 2003 & cSSTI     & 0              \\
& 203    & 268 & tigecycline                                                          & 5       & 2003 & cSSTI     & 0              \\
Stryjewski et al., 2008   & 307    & 429 & vancomycin                                                           & 6       & 2006 & cSSTI     & 0              \\
  & 309    & 426 & telavancin                                                           & 6       & 2006 & cSSTI     & 0              \\
 & 360    & 489 & vancomycin                                                           & 7       & 2006 & cSSTI     & 0              \\
  & 348    & 472 & telavancin                                                           & 7       & 2006 & cSSTI     & 0              \\
Stryjewski et al., 2006   & 81     & 95  & vancomycin                                                           & 8       & 2004 & cSSTI     & 0              \\
  & 82     & 100 & telavancin                                                           & 8       & 2004 & cSSTI     & 0              \\
Corey et al., 2010      & 297    & 347 & vancomycin                                                           & 9       & 2007 & cSSTI     & 0              \\
& 304    & 351 & ceftaroline                                                          & 9       & 2007 & cSSTI     & 0              \\
Wilcox et al., 2010     & 289    & 338 & vancomycin                                                           & 10      & 2007 & cSSTI     & 0              \\
    & 291    & 342 & ceftaroline                                                          & 10      & 2007 & cSSTI     & 0              \\
Talbot et al., 2007     & 26     & 32  & vancomycin                                                           & 11      & 2005 & cSSTI     & 0              \\
     & 59     & 67  & ceftaroline                                                          & 11      & 2005 & cSSTI     & 0              \\
Weigelt et al., 2005    & 402    & 573 & vancomycin                                                           & 12      & 2003 & cSSTI     & 0              \\
     & 439    & 583 & linezolid                                                            & 12      & 2003 & cSSTI     & 0              \\
Stevens et al., 2002    & 54     & 87  & vancomycin                                                           & 13      & 1999 & cSSTI     & 0              \\
     & 64     & 99  & linezolid                                                            & 13      & 1999 & cSSTI     & 0              \\
     & 16     & 32  & vancomycin                                                           & 14      & 1999 & pneumonia & 0              \\
     & 20     & 39  & linezolid                                                            & 14      & 1999 & pneumonia & 0              \\
Wunderink et al., 2003 & 128    & 302 & vancomycin                                                           & 15      & 2000 & pneumonia & 0              \\
  & 135    & 321 & linezolid                                                            & 15      & 2000 & pneumonia & 0              \\
Rubenstein et al., 2001  & 73     & 192 & vancomycin                                                           & 16      & 1999 & pneumonia & 0              \\
  & 85     & 203 & linezolid                                                            & 16      & 1999 & pneumonia & 0              \\
Rubenstein et al., 2011  & 221    & 374 & vancomycin                                                           & 17      & 2007 & pneumonia & 0              \\
   & 214    & 372 & telavancin                                                           & 17      & 2007 & pneumonia & 0              \\
 & 228    & 380 & vancomycin                                                           & 18      & 2007 & pneumonia & 0              \\
 & 227    & 377 & telavancin                                                           & 18      & 2007 & pneumonia & 0              \\
Fagon et al., 2000     & 67     & 148 & vancomycin                                                           & 19      & 1996 & pneumonia & 0              \\
      & 65     & 150 & \begin{tabular}[t]{c@{}l@{}}quinupristin/\\ dalfopristin\end{tabular} & 19      & 1996 & pneumonia & 0              \\
Lin et al., 2008      & NA     & 33  & linezolid                                                            & 20      & 2005 & cSSTI     & 0              \\
      & NA     & 29  & vancomycin                                                           & 20      & 2005 & cSSTI     & 0              \\
         & NA     & 38  & linezolid                                                            & 21      & 2005 & pneumonia & 0              \\
       & NA     & 40  & vancomycin                                                           & 21      & 2005 & pneumonia & 0              \\
Kohno et al., 2007     & NA     & 51  & linezolid                                                            & 22      & 2004 & cSSTI     & 0              \\
      & NA     & 26  & vancomycin                                                           & 22      & 2004 & cSSTI     & 0              \\
       & NA     & 31  & linezolid                                                            & 23      & 2004 & pneumonia & 0              \\
     & NA     & 17  & vancomycin                                                           & 23      & 2004 & pneumonia & 0              \\
Florescu et al., 2008   & NA     & 70  & tigecycline                                                          & 24      & 2005 & cSSTI     & 0              \\
 & NA     & 23  & vancomycin                                                           & 24      & 2005 & cSSTI     & 0              \\
Itani et al., 2010   & 223    & 276 & linezolid                                                            & 25      & 2007 & cSSTI     & 1              \\
     & 196    & 266 & vancomycin                                                           & 25      & 2007 & cSSTI     & 1              \\
Wunderink et al., 2008   & NA     & 30  & linezolid                                                            & 26      & 2005 & pneumonia & 1              \\
  & NA     & 20  & vancomycin                                                           & 26      & 2005 & pneumonia & 1              \\
Wunderink et al., 2012  & 102    & 186 & linezolid                                                            & 27      & 2010 & pneumonia & 1              \\
  & 92     & 205 & vancomycin                                                           & 27      & 2010 & pneumonia & 1 \\
\hline            
\end{tabular}}
\end{table}


\begin{table}[h]\scriptsize{
\begin{center}\caption{Risk ratio estimates, standard errors and 95\% confidence intervals for relative effects of antibiotics telavancin (TEL), linezolid (LIN), and mainstay therapy vancomycin (VAN)}\label{exampleres}
\begin{tabular}{llllllllllll}
\hline
              & \multicolumn{3}{c}{TEL vs VAN} &  & \multicolumn{3}{c}{LIN vs VAN} &  & \multicolumn{3}{c}{TEL vs LIN}\\
Method        & Est        & SE & 95\% CI &  & EST     & SE     & 95\% CI     \\ \cline{2-4} \cline{6-8} \cline{10-12} 
No Adjust		& 1.04  &0.028& (0.99,1.10) &  & 0.92  & 0.027 & (0.87,0.97) &&1.13&0.045&(1.05,1.22)    \\					
RE Arm        & 1.00  &0.010& (0.98,1.02) &  & 1.08  & 0.012 & (1.05,1.10)   &&0.92&0.014&(0.89,0.95) \\
G-Comp (RE)   & 1.00  &0.003&(1.00,1.00)  &  &  1.06 & 0.006 & (1.06,1.09)  &&0.94&0.005&(0.92,0.94) \\
TMLE (RE)     & 0.89  &0.106 & (0.75,1.19)  &  & 1.05 & 0.012 & (1.03,1.07)  &&0.85&0.102&(0.71,1.12)  \\
\hline   
\end{tabular}\end{center}
}\end{table}

\nocitenma{Katz:001,Arbeit:001,Florescu:001,Breedt:001,Sacchidanand:001,Stryjewski:001,Stryjewski:002,Corey:001,Wilcox:001,Talbot:001,Wunderink:001,Wunderink:002,Wunderink:003,Rubenstein:001,Rubenstein:002,Fagon:001,Itani:001}

\bibliographystylenma{DeGruyter}
\bibliographynma{netmeta2}

\end{document}